\documentclass[12pt]{article}
\usepackage{amsmath}
\usepackage{graphicx,psfrag,epsf,epstopdf}
\usepackage{enumerate}
\usepackage{natbib}
\usepackage{amsfonts}
\usepackage{epsfig,color}
\usepackage{graphicx,setspace,multirow,enumerate,url,titlesec}
\usepackage{csquotes,algorithm2e,boldline,hyperref}
\graphicspath{{Images/}}

\titleformat*{\section}{\large\bfseries}
\titleformat*{\subsection}{\normalsize\bfseries}

\newcommand{\blind}{0}

\DeclareMathOperator*{\argmin}{arg\,min}
\DeclareMathOperator*{\argmax}{arg\,max}
\RestyleAlgo{boxruled}
\newtheorem{prop}{Proposition}

\newtheorem{remark}{Remark}

\begin{document}

\def\spacingset#1{\renewcommand{\baselinestretch}%
{#1}\small\normalsize} \spacingset{1}

\if0\blind
{
  \title{\bf A Geometric Variational Approach to Bayesian Inference}
    \author{
    	Abhijoy Saha$^{1}$, Karthik Bharath$^{2}$, Sebastian Kurtek$^{1}$\\	
    	\\
    \small{$^{1}$Department of Statistics, The Ohio State University}\\
   	\small{$^{2}$School of Mathematical Sciences, University of Nottingham}\\
}
\date{}
 \maketitle
}
\fi

\if1\blind
{
  \bigskip
  \bigskip
  \bigskip
  \begin{center}
    {\LARGE\bf A Geometric Variational Approach to Bayesian Inference}
\end{center}
  \medskip
} \fi

\begin{abstract}
We propose a novel Riemannian geometric framework for variational inference in Bayesian models based on the nonparametric Fisher--Rao metric on the manifold of probability density functions. Under the square-root density representation, the manifold can be identified with the positive orthant of the unit hypersphere $S^\infty$ in $\mathbb{L}^2$, and the Fisher--Rao metric reduces to the standard $\mathbb{L}^2$ metric. Exploiting such a Riemannian structure, we formulate the task of approximating the posterior distribution as a variational problem on the hypersphere based on the $\alpha$-divergence. This provides a tighter lower bound on the marginal distribution when compared to, and a corresponding upper bound unavailable with, approaches based on the Kullback--Leibler divergence. We propose a novel gradient-based algorithm for the variational problem based on Fr\'{e}chet derivative operators motivated by the geometry of $S^\infty$, and examine its properties. Through simulations and real data applications, we demonstrate the utility of the proposed geometric framework and algorithm on several Bayesian models.
\end{abstract}

\noindent%
\vfill

\newpage
\doublespacing

	\section{Introduction}
    \label{sec:Intro}

Various algorithms based on optimization techniques, such as variational inference (VI) \citep{ghahramani1999}, variational Bayes (VB) \citep{jaakkola1997}, Black Box-$\alpha$ (BB-$\alpha$) \citep{hernandez2016} and expectation propagation (EP) \citep{minka2001}, have been successfully used to approximate the posterior distribution in the Bayesian setting. Recent advancements have made variational methods very useful for complex high-dimensional Bayesian models in view of their applicability in large scale data analysis \citep{hoffman2013, broderick2013}. In particular, VB methods have proved to be popular \citep{li2016} since they provide a lower bound on (the logarithm of) the marginal density or model evidence, thus offering a natural model selection criterion \citep{ueda2002,mcgrory2007}.

In essence, VB and Markov Chain Monte Carlo (MCMC) sampling techniques are distinct approaches to resolve the same problem of approximating the posterior distribution in a Bayesian model. In certain problems, VB methods are preferred to standard MCMC for two main reasons: MCMC suffers from high computational complexity when scaling to high dimensions, and assessing convergence of an MCMC algorithm \citep{carlin2008,cowles1996} is problematic. For a recent comparative account of the main issues with MCMC- and VB-based approaches, and for guidelines on preferring one over the other, see \cite{blei2016}.

While geometric information of the statistical model has been previously considered for improving MCMC techniques \citep{GC}, there is a striking paucity of the same in variational approaches to Bayesian inference; one exception to this is the work by \cite{babak}. The aim of our work is to demonstrate the utility in the explicit use of the intrinsic geometry of the space of probability density functions (PDFs) in variational approaches to Bayesian inference. We achieve this in two complementary ways: (1) we show how the Fisher--Rao Riemannian geometry of the space of nonparametric PDFs can be used profitably to design a parameterization-invariant variational framework; and (2) we combine the geometric framework with the use of the $\alpha$-divergence in obtaining lower and upper bounds on the marginal density for a large class of Bayesian models, noted recently as an important extension within the $\alpha$-divergence framework of \cite{li2016}.

\subsection{Background}
The inference problem is the following. For a given dataset $x$, the variational problem is to find a density $q \in \mathcal{Q}$ over the unknown, hidden parameters (or latent variables) $\boldsymbol{\theta}$ that best approximates the true posterior density $p(\boldsymbol{\theta}|x)$, by solving  $\argmin_{q \in \mathcal{Q}}\mathcal{L}(p,q)$ for a suitable distance or divergence function $\mathcal{L}$. Traditional VB methods are typified by the use of the Kullback--Leibler divergence (KLD) for $\mathcal{L}$ under the \textit{mean-field} approximation that the class $\mathcal{Q}$ consists of densities with independent marginals: $\mathcal{Q}:=\{q(\boldsymbol{\theta}) = \prod_{i} q_i(\theta_i)\}$. In conditionally conjugate models, the $q_i$s belong to the same exponential family as the complete conditional distribution $q_i(\theta_i|\boldsymbol{\theta}_{-i},x)$, where $\boldsymbol{\theta}_{-i}$ denotes all of $\boldsymbol{\theta}$ except $\theta_i$. Thus, the inference problem becomes an optimization problem of determining the distribution in the class parameterized by the natural parameter in the exponential family, which often simplifies computation. The (approximate) solution to the variational problem is usually obtained by a gradient ascent (or descent) approach along the individual coordinates of $\boldsymbol{\theta}$, where the updates are simple and available as members of the same family \citep{beal2003,bishop2006}. \cite{wang2013} extended the VB approach to nonconjugate models and proposed two generic methods which use Gaussian approximations: Laplace variational inference and delta method variational inference. More recently, there have been a few approaches in literature to relax the mean-field approximation in variational Bayes \citep{rezende2015,hoffman2015,kingma2016,kucukelbir2017}. The assumption of a specific parametric form for the class of approximating densities for the posterior, e.g., Gaussian, is a common restriction for some of the aforementioned techniques including \cite{hoffman2015} and \cite{kucukelbir2017}.


The approximating class $\mathcal{Q}$ should be large enough to include densities close to $p(\boldsymbol{\theta}|x)$. The restriction to a parametric family of distributions, e.g., exponential family, imposes restrictions on the statistical model under consideration. Moreover, the geometry of $\mathcal{Q}$ plays an important role in the performance of a gradient-based or line-search algorithm. The lack of geometric considerations of $\mathcal{Q}$ in the KLD-based VB framework was noted by \cite{hoffman2013} wherein the (approximate) natural gradient, proposed by \cite{amari1998}, capturing the curvature of the space through the Fisher information matrix, was used for updates in the gradient descent algorithm.

\subsection{Motivation and Contributions}
The proposed framework is mainly motivated by nonconjugate Bayesian models, but is equally applicable to conjugate ones, as demonstrated in the simulation examples in the sequel. Utility of VB procedures for nonconjugate models are influenced by three main inter-dependent factors: (1) choice of the variational family $\mathcal{Q}$, (2) choice of the loss function $\mathcal{L}$, and (3) computation of the gradient, and efficiency of exploration of $\mathcal{Q}$ in gradient-based algorithms. The interplay between these three factors, and their impact on the quality of posterior approximations, can be captured and quantified under a geometric framework: $\mathcal{Q}$ can be chosen in order to make it compatible with a Riemannian structure, with the resulting distance governing the choice of $\mathcal{L}$, under which local moves in the parameter space can be carried out using (Fr\'{e}chet) directional derivatives.

To this end, for a continuous parameter set of $d$ dimensions, we choose as $\mathcal{Q}$ the nonparametric manifold of \emph{all} probability densities in $d$ dimensions that factorize. We equip $\mathcal{Q}$ with the nonparametric Fisher--Rao (FR) Riemannian metric (simply referred to as the FR metric hereafter). The distinguishing feature of our approach lies in the fact that the variational problem is not defined directly on $\mathcal{Q}$, but instead on the space of all square-root PDFs. The square-root map transforms $\mathcal{Q}$ onto the positive orthant of the infinite-dimensional unit sphere in $\mathbb{L}^2$. This simplifies computations through explicit expressions for useful geometric quantities and operations (e.g., geodesic path and distance, exponential and inverse-exponential maps, parallel transport). Under such a setup, it is possible to obtain a `local linear' representation of the $d$-dimensional density as a vector in the tangent space, a subspace of a suitable Hilbert space. This allows for a representation of the density with a basis set containing an infinite number of orthornormal functions spanning the tangent space. In practice, one is required to choose a finite number of basis functions resulting in a finite-dimensional representation of the density; the theoretical framework is unencumbered by such a restriction. An upside of the truncation to an $N$-basis representation is that the choice of $N$ acts as a `bandwidth parameter' when approximating the posterior density, and can hence be tuned to improve the quality of the approximation.

Leveraging the metric structure of $\mathcal{Q}$, we use the R{\'e}nyi $\alpha$-divergence \citep{renyi1961} as the loss function $\mathcal{L}$ in the variational formulation. The $\alpha$-divergence subsumes a large family of divergences (including the KLD).  Our choice of $\mathcal{L}$ is motivated by the fact that the FR metric is closely related to the $\alpha$-divergence for $\alpha=1/2$, and also by the possibility of obtaining lower and upper bounds on the marginal density by suitably varying $\alpha$ (Section \ref{subsec:Alpha} and Proposition \ref{prop:lower_bound}), currently unavailable in existing literature. However, we note that the Riemannian framework under the FR metric can be employed using \emph{any} divergence function as a choice for $\mathcal{L}$ with appropriate adjustments. Armed with a versatile loss function $\mathcal{L}$ defined through a Riemannian metric, the gradient direction in an ascent/descent algorithm is now defined as a Fr\'{e}chet directional derivative along directions given by the orthonormal basis elements in the tangent space of the current iterate. This results in efficient explorations of $\mathcal{Q}$, as evidenced in the simulation and data analysis examples. We additionally prove the existence of an optimal step size for the gradient (Proposition \ref{stepsize}). As with any VB procedure, computing the gradient direction requires us to approximate $d$-dimensional integrals. We present a novel approximation of the gradient based on a general first-order Taylor approximation of high-dimensional integrals developed by \cite{olson1991}. Such an approximation works quite well, even in fairly high dimensions. The generality of the approximation makes it possible, in principle, to extend our framework to the non mean-field setting. We comment on this extension in Section \ref{sec:Summary} and leave it for future work.

To summarize, the main contributions of this paper are:
	\begin{itemize}
	\item We propose a Riemannian-geometric framework for variational inference for continuous $d$-dimensional densities based on the intrinsic geometry of the manifold of all PDFs equipped with the nonparametric FR metric. The approximating family $\mathcal{Q}$ contains all $d$-dimensional densities on the parameter space with independent marginals.
	\item We show, theoretically and numerically, that the proposed approach using the $\alpha$-divergence loss function results in a tighter lower bound on the marginal density than the KLD-based VB approach. Our approach is also able to provide an upper bound on the marginal, which cannot be obtained with the standard KLD-based VB.
	\item We utilize the geometry of the space of PDFs to define a gradient ascent algorithm based on Fr\'{e}chet derivatives to solve the variational problem. We also specify a technique to approximate the gradient function efficiently based on a novel first-order Taylor approximation argument.
	\end{itemize}
The rest of the paper is organized as follows. Section \ref{sec:Riem} introduces the FR Riemannian geometric framework and describes the tools relevant to our analysis.  In Section \ref{sec:VarInf}, we review the $\alpha$-divergence, and provide a detailed formulation of the variational problem within the FR framework. Further, we derive bounds for the marginal density based on an appropriate energy function closely related to the $\alpha$-divergence. In Section \ref{Algo}, we present a gradient ascent algorithm for approximating the posterior distribution, and examine its properties. In Section \ref{sec:Experimental}, we present a simulation study along with a few applications of the proposed method using various models including linear regression, density estimation and logistic regression. Section \ref{sec:Summary} includes a discussion of future work directions including possible ways to extend the proposed methodology to non-mean-field variational families.

\section{Fisher--Rao Riemannian Geometry of PDFs}
  \label{sec:Riem}

In this section, we introduce a representation space of PDFs, and associated geometric tools, which are useful in formulating the proposed variational method; most of these concepts have been previously summarized in \cite{kurtek2015,KurtekEJS}.

For simplicity, we restrict our attention to the case of univariate densities on $[0,1]$. We note however, that the framework is equally valid for all finite-dimensional distributions. Denote by $\mathcal{P}$, the Banach manifold of PDFs defined as $\mathcal{P} = \{p:[0,1] \rightarrow \mathbb{R}_{>0} \mid \int_0^1 p(x) dx = 1\}$. Next, for a point $p\in\mathcal{P}$, consider a vector space that contains the set of tangent vectors at this point. This is defined as the tangent space at the point $p$, $T_{p}(\mathcal{P}) = \{\delta p: [0,1] \rightarrow \mathbb{R} \mid \int_0^1 \delta p(x) p(x) dx = 0 \}$. Intuitively, the tangent space $T_{p}(\mathcal{P})$ at any point $p$ contains all possible perturbations of the PDF $p$. This tangent space can be used to define a suitable metric on the manifold $\mathcal{P}$ as follows. For any $p \in \mathcal{P}$ and any two tangent vectors $\delta p_1, \delta p_2 \in  T_{p}(\mathcal{P})$, the nonparametric FR metric is given by $\left\langle\langle \delta p_1, \delta p_2 \right\rangle\rangle_{p} = \int_0^1 \delta p_1(x) \delta p_2(x) \dfrac{1}{p(x)} dx$ \citep{rao1945,kass2011}. This metric is closely related to the Fisher information matrix, rendering it attractive to use in various statistical methods. An important property of this metric is that it is invariant to reparameterization \citep{cencov2000}, i.e., smooth transformations of the domain of PDFs. However, since the FR metric changes from point to point on ${\cal P}$, it leads to cumbersome computations, which makes it difficult to use in practice. Thus, instead of working on $\mathcal{P}$ directly under the FR metric, we use a suitable transformation that simplifies the Riemannian geometry of this space.

The square-root representation \citep{bhattacharyya1943} provides an elegant simplification. We define a mapping $\phi: \mathcal{P} \rightarrow \Psi$, where $\phi(p) = \psi = \sqrt{p}$ is the square-root density (SRD) of a PDF $p$; the inverse mapping is simply given by $\phi^{-1}(\psi) = p = \psi^2$ \citep{kurtek2015}. The space of all SRDs is $\Psi = \{ \psi: [0,1] \rightarrow \mathbb{R}_{>0} \mid \int_0^1 \psi^2(x) dx = 1 \}$, i.e., the positive orthant of the unit Hilbert sphere \citep{lang2012}. Since the differential geometry of the sphere is well-known, one can define standard geometric tools on this space for analyzing PDFs analytically. Let $T_\psi (\Psi) = \{ \delta \psi \mid \left\langle \delta \psi, \psi \right\rangle = 0 \}$ denote the tangent space at $\psi \in \Psi$. Under the SRD representation, it is straightforward to show that, for any two vectors $\delta\psi_1, \delta\psi_2 \in T_\psi (\Psi)$, the FR metric reduces to the standard ${\mathbb{L}^2}$ Riemannian metric: $\left\langle \delta \psi_1, \delta \psi_2 \right\rangle = \int_0^1 \delta \psi_1(t) \delta \psi_2(t) dt$. The corresponding geodesic distance between two PDFs $p_1,\ p_2 \in \mathcal{P}$, now represented by the SRDs $\psi_1,\ \psi_2 \in \Psi$, is now simply defined as the length of the shortest arc connecting them on $\Psi$: $d_{FR}(p_1,p_2) = \cos^{-1}(\left\langle \psi_1, \psi_2 \right\rangle) = \upsilon$.

We will use additional geometric tools to solve the variational inference problem in subsequent sections. These include the exponential and inverse-exponential maps, and parallel transport. For $\psi \in \Psi$ and $\delta\psi \in T_\psi (\Psi)$, the exponential map at $\psi$, $\exp_\psi: T_\psi (\Psi) \rightarrow \Psi$ is defined as $\exp_\psi(\delta\psi) = \cos(\left\| \delta\psi \right\|)\psi + \sin(\left\| \delta\psi \right\|)\dfrac{\delta\psi}{\left\| \delta\psi \right\|}$, where $\|\cdot\|$ is the ${\mathbb{L}^2}$ norm. Similarly for $\psi_1,\ \psi_2 \in \Psi$, the inverse-exponential map denoted by $\exp^{-1}_{\psi} : \Psi \rightarrow T_\psi (\Psi)$ is $\exp^{-1}_{\psi_1}(\psi_2) = \frac{\upsilon}{\sin(\upsilon)}\left( \psi_2 - \cos(\upsilon) \psi_1 \right), \quad \upsilon=d_{FR}(p_1,p_2)$. With the help of these two tools from differential geometry, we can travel between $\Psi$, the representation space of SRDs, and $T_\psi (\Psi)$. Finally, we define parallel transport, which is used to map tangent vectors from one tangent space to another. We use the parallel transport along geodesic paths (great circles) in $\Psi$. For $\psi_1,\ \psi_2 \in \Psi$, and a vector $\delta \psi \in T_{\psi_1} (\Psi)$, the parallel transport of $\delta \psi$ from $\psi_1$ to $\psi_2$ along the geodesic path is defined as $\delta \psi^{||} = \delta \psi_{\psi_1 \to \psi_2} = \delta \psi - \dfrac{2 \langle \delta \psi, \psi_2 \rangle}{\left\| \psi_1 + \psi_2 \right\|}(\psi_1 + \psi_2)$, where $\delta \psi^{||} \in T_{\psi_2} (\Psi)$. This defines a mapping $\kappa: T_{\psi_1} (\Psi) \rightarrow T_{\psi_2} (\Psi)$ such that $\delta \psi^{||} = \kappa(\delta \psi)$. An important property of parallel transport is that the mapping $\kappa$ is an isometry between two tangent spaces, i.e., for $\delta \psi_1,\ \delta \psi_2 \in T_{\psi_1} (\Psi),\ \langle \delta \psi_1, \delta \psi_2 \rangle = \langle \kappa (\delta \psi_1), \kappa (\delta \psi_2) \rangle$.
    \section{Variational Inference Based on the $\boldsymbol{\alpha}$-Divergence}
	\label{sec:VarInf}
Our objective is to synthesize the benefits of using a divergence measure that leads to lower and upper bounds on the marginal density in a Bayesian model with the Riemannian geometric structure of the space of PDFs induced by the FR metric. To this end, starting with a review of R\'{e}nyi's $\alpha$-divergence in Section \ref{subsec:Alpha}, we outline the variational problem of interest in Section \ref{subsec:Problem} and formulate the corresponding optimization problem. In Section \ref{subsec:bounds}, we show how the use of the $\alpha$-divergence provides a tighter lower bound on the marginal density compared to the standard KLD-based VB setup, and in addition, an upper bound.
\subsection{ R{\'e}nyi $\alpha$-Divergence}
    \label{subsec:Alpha}
Let us consider two probability distributions $p$ and $q$ on an $d$-dimensional set $\boldsymbol{\Theta} \subset \mathbb{R}^d$. Then, the $\alpha$-divergence $D_\alpha$ \citep{renyi1961} defined for $\{\alpha : \alpha > 0,\ \alpha \neq 1\}$ is given by $D_\alpha[p||q] = \dfrac{1}{\alpha - 1} \ln \int_{\boldsymbol{\Theta}} p(\boldsymbol{\theta})^\alpha q(\boldsymbol{\theta})^{1 - \alpha} d\boldsymbol{\theta}$. The full class of $\alpha$-divergences has the following properties: (1) $D_\alpha[p||q] \geq 0$, (2) $D_\alpha[p||q] = 0$ when $p = q$ a.e., and (3) $D_\alpha[p||q]$ is convex with respect to both $p$ and $q$. Although $D_\alpha$ can be defined for any $\alpha > 0$, certain special cases are noteworthy. In particular, $D_\alpha$ is connected to KLD in two ways: (1) $\lim_{\alpha \rightarrow 0} D_\alpha[p||q] = KL(q||p)$, and (2) $\lim_{\alpha \rightarrow 1} D_\alpha[p||q] = KL(p||q)$. These limiting cases are defined using continuity of $D_\alpha$ \citep{van2014}. With specific regard to variational inference, VB attempts to minimize $KL(q||p)$ globally, whereas EP attempts to minimize $KL(p||q)$ locally. Another special case of $D_\alpha$ is that for $\alpha=1/2$, which is very closely related to the aforementioned FR metric. In fact, this is the only choice of $\alpha$, which results in a proper distance between PDFs.
    \subsection{Problem Formulation}
    \label{subsec:Problem}
Let $x \in \mathcal{X}$ denote the observed data and $\boldsymbol{\theta} = (\theta_1, \theta_2, \dots, \theta_d) \in \boldsymbol{\Theta}$ denote the unknown $d$-dimensional parameter, where $\{\boldsymbol{\Theta} = (\Theta_1, \Theta_2, \dots, \Theta_d) : \theta_i \in \Theta_i \}$. Let $f(\boldsymbol{\theta},x) = f(x|\boldsymbol{\theta}) \pi(\boldsymbol{\theta})$ denote the joint density of $x$ and $\boldsymbol{\theta}$ where $f(x|\boldsymbol{\theta})$ is the likelihood function and $\pi(\boldsymbol{\theta})$ is the prior distribution on $\boldsymbol{\theta}$. The posterior distribution is then given by $p(\boldsymbol{\theta}|x) = \dfrac{f(x,\boldsymbol{\theta})}{m(x)}$ where $m(x) = \int_{\boldsymbol{\Theta}} f(x,\boldsymbol{\theta}) d\boldsymbol{\theta}$ denotes the marginal density of $x$, sometimes also called the model evidence. In practice, calculating the posterior is difficult because evaluating $m(x)$ is hard in general, especially when analytical solutions are not available. In such scenarios, we have to resort to approximate Bayesian inference methods as discussed in Section \ref{sec:Intro}. To this effect, we consider a variational framework based on $D_\alpha$, where we wish to find a PDF to approximate the true posterior among the class of \textit{all} joint PDFs that factorize.

Based on the mean-field approximation, let $\mathcal{Q} = \{q \mid q = \prod_{i = 1}^{d} q_i \}$ denote the class of strictly positive probability densities with support $\boldsymbol{\Theta}$ that contain independent marginals. Note that $\mathcal{Q}$ is an infinite-dimensional set of PDFs on $\boldsymbol{\Theta}$, and not a parametric class. Then, the $\alpha$-divergence between the posterior and an element of $\mathcal{Q}$ is
\[
D_\alpha[p||\prod_{i = 1}^d q_i] = \dfrac{1}{\alpha - 1} \ln \int_{\boldsymbol{\Theta}} {p(\boldsymbol{\theta}|x)}^\alpha \Big( {\prod_{i = 1}^d q_i(\theta_i)} \Big)^{1 - \alpha} d\boldsymbol{\theta}, \quad \alpha > 0.
\]
Note that for the limiting case of $\alpha \to 1$, $D_\alpha$ converges to the KLD between $p$ and $q$, i.e., $\int_{\boldsymbol{\Theta}} \ln \Big( \dfrac{p(\boldsymbol{\theta}|x)}{q(\boldsymbol{\theta})} \Big) p(\boldsymbol{\theta}|x) d\boldsymbol{\theta}$. Since the integral in this case is with respect to the computationally intractable posterior density $p$, the optimization problem becomes difficult to handle. Thus, we do not consider this limiting case in our setup.

Minimizing $D_\alpha$ over $\mathcal{Q}$ is not straightforward for two reasons: (1) the nonlinear manifold structure of $\mathcal{Q}$; (2) unavailability of analytical expressions for corresponding geometric quantities. In order to exploit the FR geometry of the space of probability densities for the task of minimizing $D_\alpha$, we use the SRD representation defined in Section \ref{sec:Riem}. Accordingly, the set $\mathcal{Q}_\psi = \{\psi_q \mid \psi_q = \prod_{i=1}^{d} \psi_{q_i} \}$ consists of elements of the $d$-fold product space $\Psi_d = \Psi \times \Psi \times \dots \times \Psi$ of SRDs. Suppose the SRDs of the joint, marginal and the posterior are denoted by $\psi_f,\ \psi_m$ and $\psi_p$, respectively, and observe the following equivalence relationships:
	\begin{align*}
	(q_1^*,\ q_2^*, \dots,\ q_d^*) &= \argmin_{\mathcal{Q}}  D_\alpha[p||\prod_{i = 1}^d q_i]
    = \argmin_{\Psi_d} \dfrac{1}{\alpha - 1} \int_{\boldsymbol{\Theta}} {\psi_p(\boldsymbol{\theta}|x)}^{2\alpha} ({\prod_{i = 1}^d \psi_{q_i}(\theta_i)})^{2 - 2\alpha} d\boldsymbol{\theta} \\
    &= \argmin_{\Psi_d} \dfrac{1}{\alpha - 1} \int_{\boldsymbol{\Theta}} {\psi_f(x,\boldsymbol{\theta})}^{2\alpha} ({\prod_{i = 1}^d \psi_{q_i}(\theta_i)})^{2 - 2\alpha} d\boldsymbol{\theta}.
	\end{align*}
The last equality follows from the fact that $\psi_m(x)$, the SRD of the marginal $m(x)$, is constant in $\boldsymbol{\theta}$. Furthermore, when $\alpha < 1$, the factor $(\alpha - 1)^{-1} < 0$, and thus the minimization problem can be written as one of maximization. We can hence transfer the variational problem defined on the manifold $\mathcal{Q}$ of PDFs on $\boldsymbol{\Theta}$ to the $d$-fold product space $\Psi_d$ of SRDs whose geometry is well-understood.
Consequently, we define the energy functional $\mathcal{E}_\alpha(\psi_q; \boldsymbol{\theta}):\Psi_d \to \mathbb{R}_{>0}$ for a given element $\psi_q=\prod_{i=1}^d\psi_{q_i}$ of $\Psi_d$ as
\[
\mathcal{E}_\alpha (\psi_q;\boldsymbol{\theta}): = \int_{\boldsymbol{\Theta}} {\psi_f(x,\boldsymbol{\theta})}^{2\alpha} ({\prod_{i = 1}^d \psi_{q_i}(\theta_i)})^{2 - 2\alpha} d\boldsymbol{\theta}.\]

The case $\alpha=1/2$, as mentioned earlier, links to the intrinsic FR Riemannian metric on the space of probability densities, and is therefore coordinate-invariant. A convenient byproduct of this is that the energy functional enjoys a certain invariance.
\begin{prop}
\label{prop:invariance}
 Consider injective, differentiable coordinate reparametrizations $\phi_i:\Theta_i \to \Theta_i$ such that $\eta_i=\phi_i(\theta_i)$ for $i=1,\ldots,d$ and $\boldsymbol{\eta}=(\eta_1,\ldots,\eta_d)$. The energy functional $\mathcal{E}_{1/2}(\psi_q;\cdot)$ satisfies the invariance property
 \[
 \mathcal{E}_{1/2}(\psi_q;\boldsymbol{\eta})= \mathcal{E}_{1/2}(\psi_q;\boldsymbol{\theta}).
 \]
 \end{prop}

\begin{remark}
For simplicity, the coordinate reparameterizations $\phi$ were defined as self-maps of $\Theta_i$. Indeed, the $\phi_i$ can map $\Theta_i$ to another space altogether, but the result of Proposition \ref{prop:invariance} would still hold as long as $\phi_i$ is injective and differentiable for each $i=1,\ldots,d$. Importantly, it is easy to see that Proposition \ref{prop:invariance} holds only for $\alpha=1/2$ when integrating with respect to Lebesgue measure; the result does not hold for general reparameterizations $\eta_i=\phi_i(\theta_1,\ldots,\theta_d),\ i=1,\dots,d$ since the Jacobian matrix is no longer diagonal and the corresponding determinant of the Jacobian cannot be expressed as a product of differentials.
\end{remark}
For a general $\alpha>0$, we define the variational problem for approximating the posterior as
\[\argmax_{\Psi_d} \mathcal{E}_\alpha(\psi_q; \cdot)\text{ if }\alpha \in (0, 1)\ \ \text{or}\ \ \argmin_{\Psi_d} \mathcal{E}_\alpha(\psi_q; \cdot)\text{ if }\alpha \in (1,\infty).
\]
The definition of the energy functional $\mathcal{E}_\alpha$ distinguishes our approach to alternative variational formulations on the space $\mathcal{Q}$ under a class of distance or divergence measures: in our setup, the variational problem is defined on $\Psi_d$, and we explicitly incorporate and utilize the underlying geometry of $\Psi_d$ in minimizing $\mathcal{E}_\alpha$.

\subsection{Bounds on the Marginal Density}
  \label{subsec:bounds}
The two important reasons for using $D_\alpha$ (and not necessarily $D_{1/2}$ or KLD) are:
\begin{enumerate}
\item It leads to a tighter lower bound on the marginal density than KLD.
\item It leads to an upper bound on the marginal density, which is not possible under KLD.
\end{enumerate}
Recall that, under the traditional KLD-based VB setup, one minimizes the KLD between a member of the approximating class $q$ and the true posterior $p$:
	\begin{align*}
    (q_1^{*KL},\ q_2^{*KL}, \dots,\ q_d^{*KL}) & = \argmin_{q \in \mathcal{Q}}  \int_{\boldsymbol{\Theta}} \ln \Big( \dfrac{q(\boldsymbol{\theta})}{p(\boldsymbol{\theta}|x)} \Big) q(\boldsymbol{\theta}) d\boldsymbol{\theta}
    = \argmax_{q \in \mathcal{Q}}  \int_{\boldsymbol{\Theta}} \ln \Big( \dfrac{p(\boldsymbol{\theta}|x)} {q(\boldsymbol{\theta})} \Big) q(\boldsymbol{\theta}) d\boldsymbol{\theta} \\
    &= \argmax_{q \in \mathcal{Q}}  \int_{\boldsymbol{\Theta}} \ln \Big( \dfrac{f(x,\boldsymbol{\theta})} {q(\boldsymbol{\theta})} \Big) q(\boldsymbol{\theta}) d\boldsymbol{\theta}
    =: \argmax_{q \in \mathcal{Q}} \mathcal{H}(f,q),
	\end{align*}
where the third equality again stems from the fact that the marginal does not depend on $\boldsymbol{\theta}$. Thus, instead of minimizing $KL (q||p)$, one can choose to maximize $\mathcal{H}(f,q)$ to obtain an equivalent solution to the original optimization problem.

For a general variational family (not necessarily one which factorizes), we formally state the two results given earlier on the logarithmic scale for ease of comparison with the KLD-based bound on the marginal density.
    \begin{prop}
    \label{prop:lower_bound}
    The following inequalities hold for the marginal $m(x)$:
    \begin{enumerate}[(i)]
    \item For $0 < \alpha < 1: \mathcal{H}(f,q) \leq \dfrac{1}{\alpha} \ln \mathcal{E}_\alpha(\psi_q; \cdot) \leq \ln m(x)$, i.e., $D_\alpha$ provides a tighter lower bound on the marginal than KLD.
    \item For $\alpha > 1: \ln m(x) \leq \dfrac{1}{\alpha} \ln \mathcal{E}_\alpha(\psi_q; \cdot)$, i.e., $D_\alpha$ provides an upper bound on the marginal.
    \end{enumerate}
    \end{prop}
This proposition motivates the study of the properties of variational inference based on $D_\alpha$. In addition, the ability to compute a tighter lower bound and an upper bound on the marginal provides a novel approach to approximate Bayesian statistical inference. For example, we are able to bound the Bayes factor (ratio of two marginal densities under two models) above and below, providing better evidence for model choice.

    \section{Optimization via Gradient Ascent}

    \label{Algo}
    The definition of the energy functional $\mathcal{E}_\alpha:\Psi_d \to \mathbb{R}_{>0}$ does not require the geometric tools or the novel representation space of PDFs defined in Section \ref{sec:Riem}. Indeed, the minimum of $\mathcal{E}_\alpha$ on $\Psi_d$ is independent of the Riemannian metric and the corresponding geometric tools. However, its determination through a line-search algorithm based on gradients of $\mathcal{E}_\alpha$ is inextricably linked to the geometry of $\Psi_d$ through the Fr\'{e}chet or directional derivatives. Without restricting the class of approximating densities to parametric families, we will utilize Riemannian optimization tools under the FR framework and propose a gradient-based algorithm. Throughout this section, the subscript $i=1,\ldots,d$ indexes quantities related to the parameter $\theta_i$.

The tangent space $T_{\psi_{q_i}}(\Psi)=\{\delta \psi_{q_i}:\Theta_i \to \mathbb{R} \mid \int_{\Theta_i}\delta \psi_{q_i}(\theta_i)\psi_{q_i}(\theta_i)d\theta_i=0\}$ at $\psi_{q_i} \in \Psi$ is the vector subspace of square-integrable functions from $\Theta_i $ to $\mathbb{R}$. This space is spanned by the set $\mathcal{B}_i=\{b^k_i,\ k=1,2,\ldots\}$ of orthonormal basis functions such that $\int_{\Theta_i}b^k_i(\theta_i)\psi_{q_i}(\theta_i)d\theta_i=0\ \ \forall\ \ k$. The mean-field approximation on the class $\Psi_d$ ensures that the gradient of $\mathcal{E}_\alpha$ can be computed for its restriction $\mathcal{E}_{\alpha|\Psi}:\Psi\to \mathbb{R}_{>0}$ to $\Psi$ for each $i=1,\ldots,d$.
The Hilbert space structure of the tangent space plays a crucial role in this computation.
\begin{prop}
\label{gradient}
For each $i=1,\ldots,d$, the gradient $\nabla \mathcal{E}^i_{\alpha}$ along direction $b^k_i$ is given by:
\[
\nabla \mathcal{E}_\alpha^i=\sum_{k=1}^\infty D^i\mathcal{E}_\alpha (b^k_i) b^k_i=2(1-\alpha) \sum_{k=1}^{\infty} \Bigg[ \int_{\boldsymbol{\Theta}} \psi_{f} (x,\boldsymbol{\theta})^{2\alpha} \prod_{j \neq i} \psi_{q_j}(\theta_j)^{2 - 2\alpha} \psi_{q_i}(\theta_i)^{1 - 2\alpha} b^k_i(\theta_i) d\boldsymbol{\theta} \Bigg] b^k_{i}.
\]
\end{prop}
\begin{remark}
The gradient $\nabla \mathcal{E}_\alpha^i$ represents an ascent or a descent direction depending on whether $\alpha$ is lesser or greater than one, respectively. To unify the two cases, we use $|\nabla \mathcal{E}_\alpha^i|$ to denote the value of the map $\Theta_i\ni\theta_i\mapsto|\nabla \mathcal{E}_\alpha^i(b^k_i(\theta_i))| \in \mathbb{R}_{>0}$ at a fixed $\theta_i$. This ensures that the gradient always represents an ascent direction regardless of the value of $\alpha$.
\end{remark}
We use the geometry of the space $\Psi$ to define an appropriate basis set $\mathcal{B}_i,\ i=1,\ldots,d$. We explain the construction of this basis for $\theta_i \in [0,1]$ and note that it is easily extended to a general compact support. For this purpose, we use the tangent space at the SRD of the uniform distribution $u_i$ on $[0,1]$ defined as $T_{\psi_{u_i}}(\Psi) = \{ \delta \psi_{u_i}:[0,1] \rightarrow \mathbb{R} \mid \int_0^1 \delta \psi_{u_i} (\theta_i) d\theta_i = 0 \}$. We define the basis set $\mathcal{\tilde{B}}_i = \{ \sin(2 \pi n \theta_i), \cos(2 \pi n \theta_i), 1 - \theta_i \mid n \in \mathbb{Z}_{+}\}$. It is easy to verify that all elements of this set are orthogonal to $\psi_{u_i}$. This basis is then orthonormalized using the Gram-Schmidt procedure under the $\mathbb{L}^2$ metric to result in $\mathcal{B}_i$.

The above construction leads to an orthonormal basis only for $T_{\psi_{u_i}}(\Psi)$; it can be extended to every point of $\Psi$ using parallel transport (Section \ref{sec:Riem}). The explicit expressions for parallel transport ensure that this can be done exactly, and that the resulting basis elements in the tangent space of the new point are orthonormal and remain orthogonal to the representation space. For implementing the algorithm practically, we need to choose a finite basis set. We let $N$ denote the number of basis functions. This leads to the following gradient ascent algorithm for optimizing $\mathcal{E}_\alpha$ on $\Psi_d$ (Algorithm \ref{alg:gradasc}).

 \begin{algorithm}
 \setstretch{1.2}
Initialize: $\psi^0_{q_1}=\sqrt{q_1^0},\ldots,\psi^0_{q_d}=\sqrt{q_d^0}$,\ $l=0,\ \delta>0$, $l_{max}$ and $N$\;
For each $i=1,\ldots,d$, select orthonormal bases $\mathcal{B}_i=\{b_i^k:\Theta_i \to \mathbb{R},\ k=1,\ldots,N\}$ for tangent spaces $T_{\psi^0_{q_i}}(\Psi)$\;
Choose step size $\epsilon>0$\;
\While{$\min \{\nabla \mathcal{E}^i_\alpha,\ i=1\ldots,d\} > \delta$ \textbf{and} $l < l_{max}$ }
{
Compute ascent direction
$|\nabla \mathcal{E}_\alpha^i|$\;
Update: $\psi^{l+1}_{q_i}=\text{exp}_{\psi^{l}_{q_i}}\left(\epsilon |\nabla \mathcal{E}^i_\alpha|\right)$\;
Parallel transport basis: $b^k_i=b^k_{\psi^0_{q_i} \to \psi^{l+1}_{q_i}}$ for $i=1,\ldots,d$ and $k=1,\ldots,N$\;
$l = l + 1$\;
}
Return $\psi^l_{q_i},\ i=1,\ldots,d$.
 \caption{Gradient-ascent algorithm on $\Psi_d$.}\label{alg:gradasc}
 \end{algorithm}

A key aspect of the algorithm is the availability of an explicit expression for the exponential map, which ensures that we remain in the space of SRDs. Our approach is then to separately update each $\psi_{q_i}$ at every iteration until convergence. As this is a gradient-based approach, we are not guaranteed to arrive at the global solution. There are many approaches to initialize the algorithm. However, through simulation, we found that initialization does not play a crucial role with respect to convergence. In related work, \cite{minka2005} defined optimization algorithms for $D_\alpha$, but under the assumption that the approximating class is an exponential family. The proposed geometric approach is more general.

\subsection{Choice of Step Size and Approximation of the Gradient}
    \label{subsec:Estimation}

The performance of the algorithm on $\Psi_d=\Psi \times \cdots \times \Psi$ is governed by its performance on the individual $\Psi$. The computation of the gradient $\nabla \mathcal{E}^i_\alpha$ and the choice of the step size $\epsilon$ are crucial in order for the algorithm to efficiently explore $\Psi$. For finite-dimensional optimization problems, the existence of an optimal $\epsilon$ that guides its selection is given by the so-called Wolfe-conditions.

The proposed ascent algorithm is defined on an infinite-dimensional manifold; the corresponding Wolfe-conditions can be defined in terms of the functional $\tilde{\mathcal{E}}^{i}_\alpha: T_{\psi_{q_i}}(\Psi) \to \mathbb{R}_{>0}$ with $\tilde{\mathcal{E}}^{i}_\alpha(v_i)=\mathcal{E}^{i}_\alpha \circ \text{exp}(v_i)$ for a tangent vector $v_i$. Note that $\tilde{\mathcal{E}}^{i}_\alpha$ is now an element of the dual space of $T_{\psi_{q_i}}(\Psi)$, which is a linear subspace of $\mathbb{L}^2(\Theta_i)$. For a given ascent direction $v_i \in T_{\psi_{q_i}}(\Psi)$, the corresponding (weak) Wolfe-conditions that specify guidelines for the choice of the step size $\epsilon$ are given by \citep{RW}:
\begin{align}
\label{wolfe}
\tilde{\mathcal{E}}^{i}_\alpha(\epsilon v_i)&\geq \mathcal{E}^{i}_\alpha(\psi_{q_i})+c_1\epsilon D^i\mathcal{E}_\alpha(v_i)\ \ \text{and}\ \ D^i\tilde{\mathcal{E}}^i_\alpha(\epsilon v_i)D^i\text{exp}(\epsilon v_i) v_i \leq c_2D^i\mathcal{E}_\alpha(v_i),
\end{align}
where $D^i\text{exp}(\epsilon v_i)$ is the derivative of the exponential map at $v_i \in T_{\psi_{q_i}}(\Psi)$, $D^i\tilde{\mathcal{E}}^i_\alpha(\epsilon v_i)$ is the directional derivative of $\tilde{\mathcal{E}}_{\alpha|\Psi}$, the restriction of $\tilde{\mathcal{E}}_{\alpha}$ to $\Psi$, and $0<c_1<c_2<1$. It does not follow directly that, for a given algorithm on the infinite-dimensional manifold, an $\epsilon$ satisfying Equation \ref{wolfe} exists. The following result clarifies this for the proposed approach.
\begin{prop}
\label{stepsize}
For an ascent direction $v_i \in T_{\psi_{q_i}}(\Psi)$, an $\epsilon$ satisfying the Wolfe conditions in Equation \ref{wolfe} exists.
 \end{prop}

One significant issue encountered when computing the gradient is the evaluation of an integral over the $d$-dimensional $\boldsymbol{\Theta}$. While the mean-field approximation on $\psi_q$ helps, the presence of the (square-root) joint density $\psi_f(x,\boldsymbol{\theta})$ in the integrand complicates matters. We use a nested univariate first-order Taylor approximation of the multivariate integral proposed in \cite{olson1991}, which reduces a multivariate integral to functions of univariate ones. Briefly, the basis of the approximation method is as follows. Let $y$ be a random variable with $E(y) = \mu$. Suppose we are interested in evaluating $E(g(y))$ for a smooth function $g$. The first-order Taylor expansion of $g$ around $\mu$ is $g(y) = g(\mu) + g'(\mu) (y - \mu) + O_p(y - \mu)^2$. Taking expectations on both sides, we obtain $E(g(y)) = g(\mu) + 0 + O(V(y))$. Thus, $E(g(y))$ is approximated with $g(\mu)$.

For $d$-dimensional $\boldsymbol{\theta}$, consider the approximation of $E(g(\boldsymbol{\theta})) = \int_{\boldsymbol{\Theta}} g(\boldsymbol{\theta}) f(\boldsymbol{\theta}) d\boldsymbol{\theta}$. Using the above argument, $E(g(\boldsymbol{\theta}))$ can be expressed as:
	\begin{align*}
	E(g(\boldsymbol{\theta})) &= \int_{\theta_d} \Bigg[ \int_{\theta_1 \times \theta_2 \times \dots \times \theta_{d-1}} g(\theta_1, \theta_2, \dots, \theta_d) f(\theta_1, \theta_2, \dots, \theta_{d-1} | \theta_d) d\theta_1 d\theta_2 \dots d\theta_{d-1} \Bigg] f(\theta_d) d\theta_d \nonumber \\
	&= E_{\theta_d} \Bigg[ \int_{\theta_1 \times \theta_2 \times \dots \times \theta_{d-1}} g(\theta_1, \theta_2, \dots, \theta_d) f(\theta_1, \theta_2, \dots, \theta_{d-1} | \theta_d) d\theta_1 d\theta_2 \dots d\theta_{d-1} \Bigg],
	\end{align*}
	where $f(\theta_1, \theta_2, \dots, \theta_{d-1} | \theta_d)$ is the density of $(\theta_1, \theta_2, \dots, \theta_{d-1})$ conditional on $\theta_d$, and $E_{\theta_d}$ denotes expectation with respect to $\theta_d$. Let $\mu_d = E_{\theta_d}(\theta_d) = \int_{\theta_d} \theta_d f(\theta_d) d\theta_d$. We use a first-order Taylor expansion to approximate the conditional expectation above about $\mu_d$: $E(g(\boldsymbol{\theta})) \approx \int_{\theta_1 \times \theta_2 \times \dots \times \theta_{d-1}} g(\theta_1, \theta_2, \dots, \theta_{d-1}, \mu_d) f(\theta_1, \theta_2, \dots, \theta_{d-1} | \mu_d) d\theta_1 d\theta_2 \dots d\theta_{d-1}$. We can keep on repeating the above approximation technique until we obtain the univariate integral $E(g(\boldsymbol{\theta})) \approx \int_{\theta_1} g(\theta_1, \mu_{2|3,\dots,d}, \mu_{3|4,\dots,d},\dots,\mu_d) f(\theta_1 | \mu_{2|3,\dots,d}, \mu_{3|4,\dots,d},\dots,\mu_d) d\theta_1$,
where $\mu_{j | j+1,\dots,d}$ is the conditional expectation of $\theta_j | \theta_{j+1}, \dots, \theta_d$, $j = 2, \dots, d-1$.
	
Consider the expression for the gradient $\nabla \mathcal{E}^i_\alpha$.  Bearing in mind that in our setting the joint density is $q=\prod_j q_j$, applying the above approximation we can rewrite the integral in the expression for the gradient as
	\begin{equation}
		\label{eq:integral}
        \int_{\Theta_{d}} \dots \int_{\Theta_{2}} \int_{\Theta_{1}} \dfrac{\psi_f(x, \theta_1, \theta_2, \dots ,\theta_d)^{2\alpha}}{\prod_{j \neq i} \psi_{q_j} (\theta_j)^{2\alpha}} \prod_{j \neq i} q_j(\theta_j) \psi_{q_i}(\theta_i)^{1 - 2\alpha} b_i^{k}(\theta_i) d\theta_1 d\theta_2 \dots d\theta_d,
	\end{equation}
since $q_j(\theta_j) = \psi_{q_j}^2(\theta_j)$. We first compute the expectations $\mu_j = \int_{\Theta_j} \theta_j q_j(\theta_j) d\theta_j,\ \forall\ j \neq i$. We then use these expected values to redefine the high dimensional integral as a one dimensional integral given by
\[
\int_{\Theta_i} \dfrac{\psi_f (x, \mu_{-i}, \theta_i)^{2\alpha}}{\prod_{j \neq i} \psi_{q_j} (\mu_j)^{2\alpha}}\psi_{q_i}(\theta_i)^{1 - 2\alpha} b_i^{k}(\theta_i) d\theta_i,
\]
where $\mu_{-i}$ denotes all of $\boldsymbol{\mu}$ except $\mu_i$. We apply the same first-order Taylor expansion technique to approximate the bounds on the marginal density as defined in Proposition \ref{prop:lower_bound}.
\section{Simulations and Real Data Examples}
  \label{sec:Experimental}

In this section, we present several examples that validate the proposed framework. In the first example, we consider a simulation study from a normal-gamma conjugate model where the posterior distribution is bivariate. Since the true value of the marginal density is known in this case, we can compare the marginal for a given dataset $x$ to the bounds computed under our setup and to the lower bound obtained using KLD. Next, we assess the performance of our method in the context of Bayesian multiple linear regression and Bayesian density estimation using logistic Gaussian process priors. The last model we consider is logistic regression. In this case, we compare classification performance of our method to various other techniques. Finally, we consider a real signature verification experiment using novel shape-based signature descriptors.

    \subsection{Low-Dimensional Simulation Study}
    \label{subsec:Simulation}
We consider the following hierarchical model: $x|\mu,\tau \overset{iid}{\sim} N(\mu,\tau^{-1}),\ \ \ \mu|\tau \sim N(0, \tau^{-1}),\ \ \ \tau \sim Ga(0.01, 0.01).$
Because the posterior in this case is bivariate, we can evaluate the proposed method using the ``ground truth''. Additionally, we can compare the estimated marginal computed using our method and that computed under KLD. As described earlier, based on the mean-field approximation, we assume that the posterior distribution factorizes: $q(\mu, \tau) = q(\mu)q(\tau)$. It is easy to show that under the KLD-based VB, the optimal distribution of $\mu$ is $q^{*KL}(\mu) = N(\mu_0^*,\lambda_0^{*-1})$, and the optimal distribution of $\tau$ is $q^{*KL}(\tau) = Ga(a^*,b^*)$. Thus, only the parameters of these two distributions need to be updated at each iteration. The updates are given by $\mu_0^* = \dfrac{n\bar{x}}{1+n},\ \lambda_0^* = (1+n) \int_{\mathbb{R}_{>0}}\tau q(\tau) d\tau$, $a^* = 0.01 + \dfrac{n+1}{2}$ and $b^* = 0.01 + \dfrac{1}{2}\int_\mathbb{R}(2\mu^2 + (\sum_{i=1}^n x_i)^2 -2\mu\sum_{i=1}^n x_i)q(\mu) d\mu$, where $n$ is the sample size and $\bar{x}$ is the sample mean In the proposed algorithm, we use only 99 basis elements to show the efficiency of our method. Multiple simulation studies reveal that increasing the number of basis elements can lead to better approximations of the posterior.

	\begin{figure}[!t]
		\begin{center}
			\begin{tabular}{|c|c|c|}
\hline
(a) $x_i \overset{iid}{\sim} N(0,1), i=1,\dots,100$ &
(b) $x_i \overset{iid}{\sim} U(0,2), i=1,\dots,100$&
(c) $x_i \overset{iid}{\sim} t_2, i=1,\dots,20$\\
\hline
\includegraphics[width=1.4in]{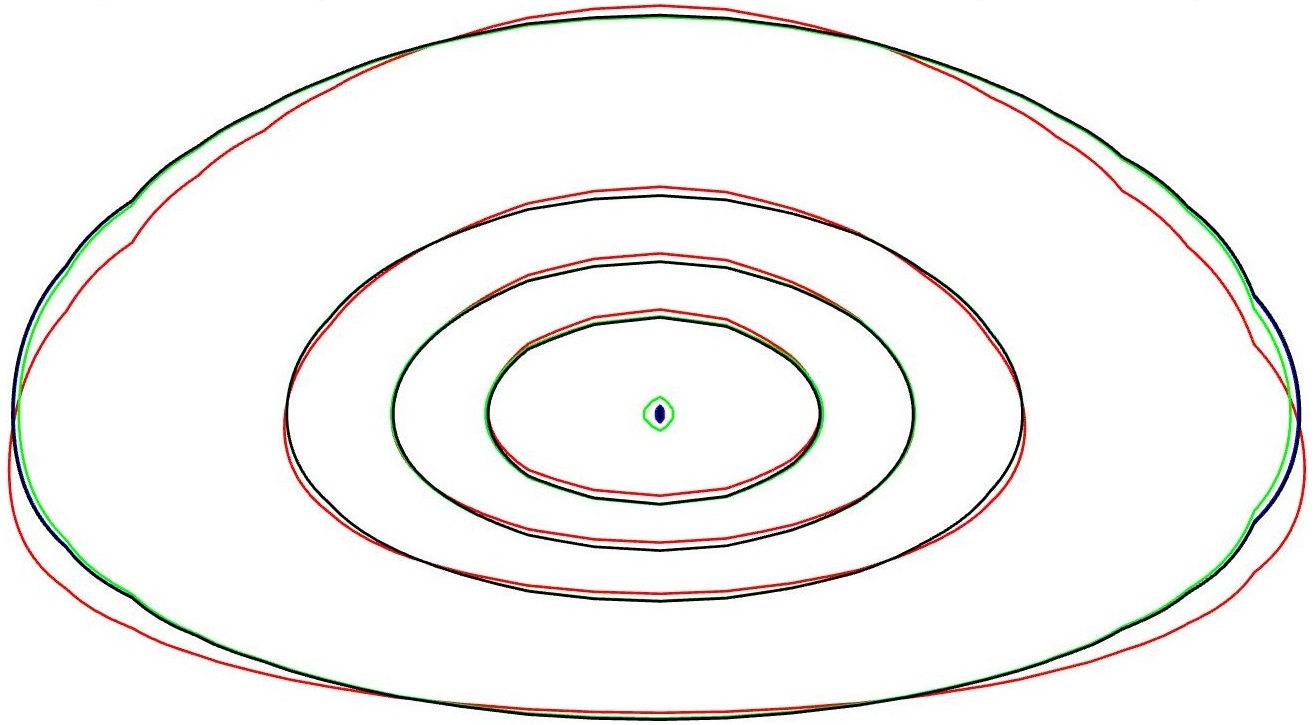}&\includegraphics[width=.5in]{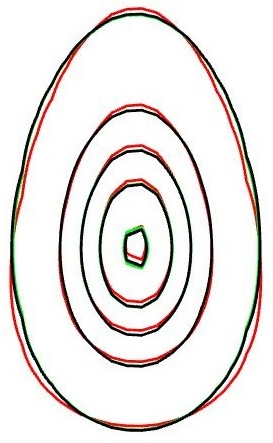}&\includegraphics[width=1.4in]{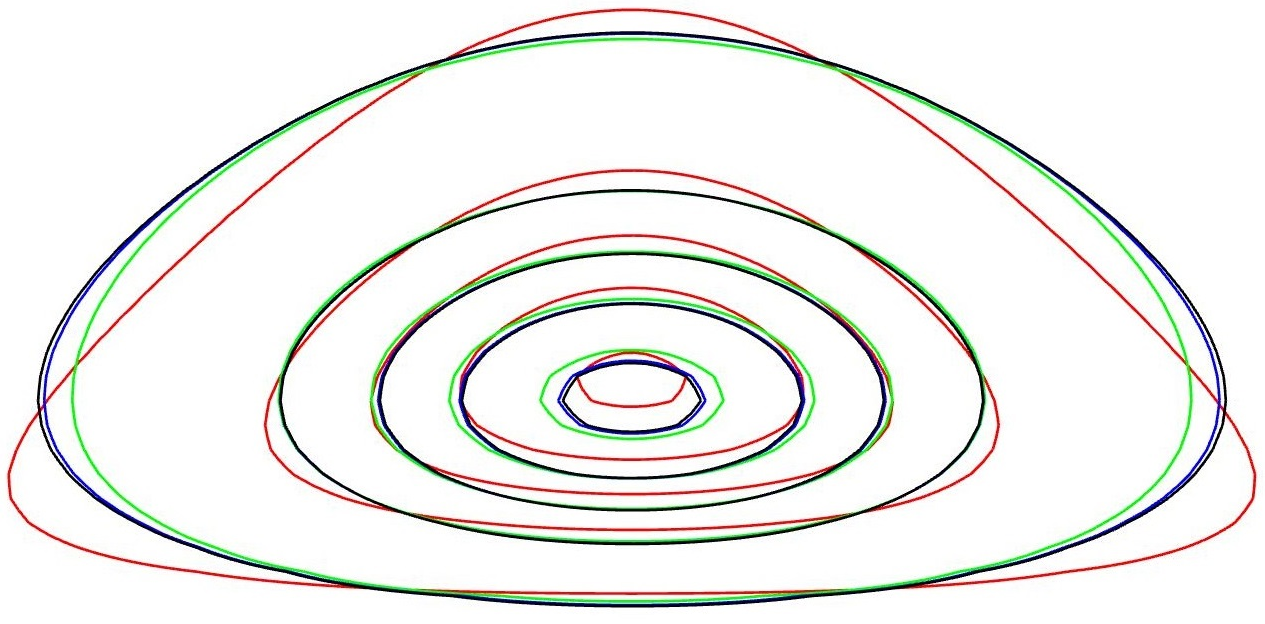}\\
\hline
$LB_{PM}=0.9995$&$LB_{PM}=0.9995$&$LB_{PM}=0.9975$\\
$UB_{PM}=1.0005$&$UB_{PM}=1.0005$&$UB_{PM}=1.0025$\\
$LB_{KLD}=0.9950$&$LB_{KLD}=0.9950$&$LB_{KLD}=0.9757$\\
\hline
\hline
\includegraphics[width=1.4in]{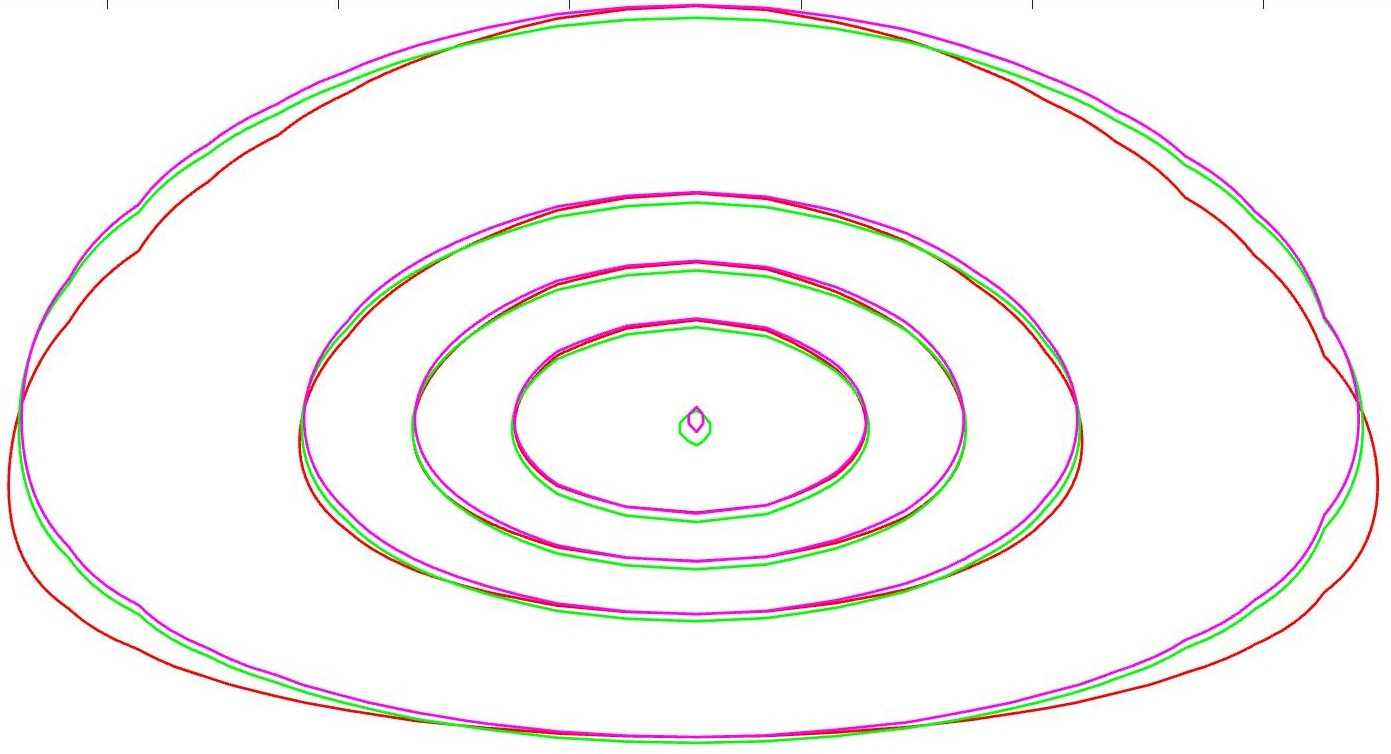}&\includegraphics[width=.5in]{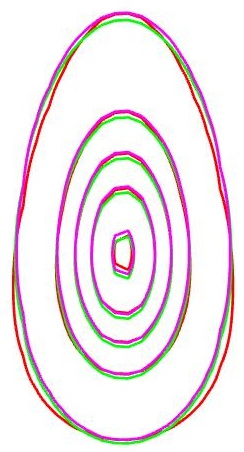}&\includegraphics[width=1.4in]{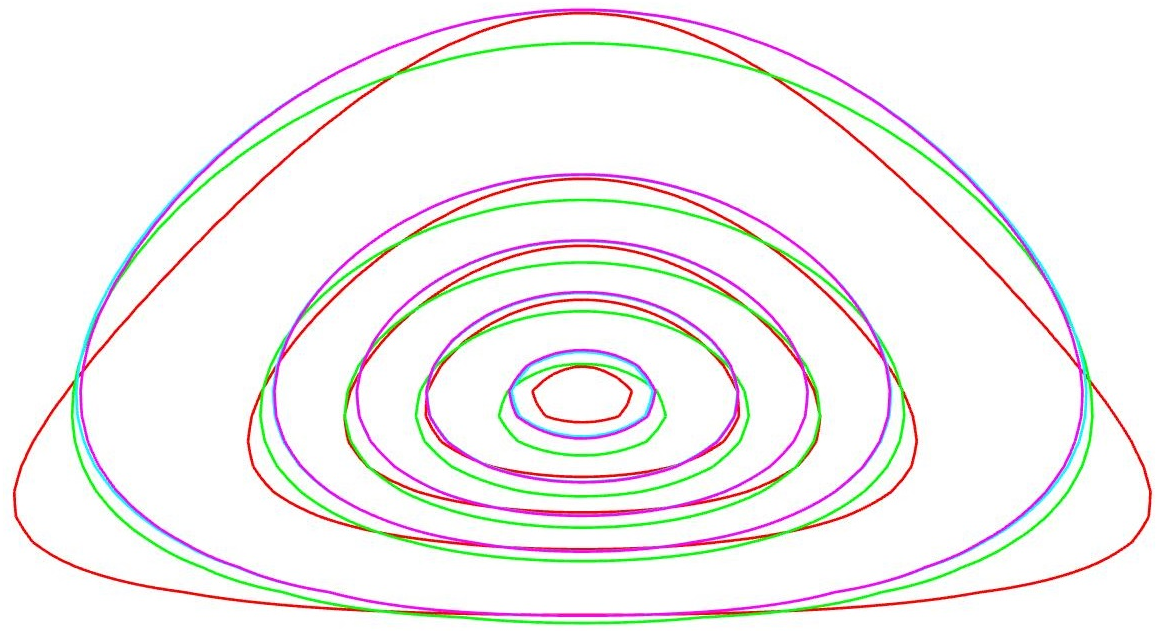}\\
\hline
$LB_{PMA}=0.9992$&$LB_{PMA}=0.9992$&$LB_{PMA}=0.9960$\\
$UB_{PMA}=1.0008$&$UB_{PMA}=1.0008$&$UB_{PMA}=1.0042$\\
$LB_{KLD}=0.9950$&$LB_{KLD}=0.9950$&$LB_{KLD}=0.9757$\\
\hline
\end{tabular}
		\end{center}
		\caption{Contour plots of the approximated posteriors and the true posterior for three different simulated datasets. LB = lower bound, UB = upper bound, PM = proposed method, KLD = Kullback-Leibler divergence and PMA = proposed method with approximated integral. All of the values are to be compared to the optimal value of 1.}
		\label{fig:simu}
	\end{figure}

We compare three different approaches: KLD-based VB (KLD), the proposed method with the gradient evaluated using a numerical integral (PM), and the proposed method with the gradient evaluated using the approximation described in Section \ref{subsec:Estimation} (PMA). Using Proposition \ref{prop:lower_bound}, the lower ($LB_{PM}$, $LB_{KLD}$) and upper ($UB_{PM}$) bounds on the marginal can be computed exactly in this scenario, since it only involves a two-dimensional integral. To show the efficiency of the proposed first-order integral approximation technique in this low-dimensional study, we calculate the lower ($LB_{PMA}$) and upper ($UB_{PMA}$) bounds for our method using the approximation described in Section \ref{subsec:Estimation} as well. The evaluation is done on three simulated datasets as shown in Figure \ref{fig:simu}. For each of the simulations, we use $\alpha = 0.9$ for the lower bound (LB) and $\alpha = 1.1$ for the upper bound (UB) on the marginal. Figure \ref{fig:simu} displays the comparison of contour plots of the true posterior and other posterior approximations using the techniques discussed above. For all images, the true posterior is plotted in red and the KLD solution is plotted in green. The top row contains the posterior approximations based on the proposed method without the integral approximation, where $LB_{PM}$ and $UB_{PM}$ are plotted in blue and black, respectively. The bottom row contains the same results computed with the integral approximation, where $LB_{PMA}$ and $UB_{PMA}$ are plotted in cyan and magenta, respectively.

For improved presentation and for ease of comparison across different simulations, we rescale the bound values such that the optimal value is 1. In all cases, the different posterior approximations are very close to the true posterior, especially when the sample size is high. We also note that the LB on the marginal computed using PM and PMA is always tighter than the KLD one. Furthermore, the main advantage of PM/PMA is that it can also compute an UB on the marginal. Panel (c) shows that the proposed method is better at estimating the tails of the posterior than KLD. Table \ref{tab:BayesFactor} shows the utility of the proposed method in statistical inference. Here, we use the first dataset (Figure \ref{fig:simu}(a)). First, we report the LB and UB on the posterior mean of both parameters $\mu$ and $\tau$. Second, we compute the LB and UB for the Bayes factor where Model (1) uses a $N(0, \tau^{-1})$ prior, and Model (2) uses a $N(2, \tau^{-1})$ prior. We note that the bounds on the posterior means and Bayes factor are very tight. In fact, the difference between the bounds is smaller than $1 \times 10^{-5}$ in the posterior mean case. Furthermore, the Bayes factor suggests that Model (1) (prior mean is $0$) is better than Model (2), which is in line with our expectation (since the data was sampled from a $N(0,1)$). These results suggest that the proposed approach has promise when extended to higher-dimensional and more complex Bayesian models.

        	\begin{table}[!t]
\centering
\begin{tabular}{|cc|cc|cc|}
\hline
\multicolumn{2}{|c|}{Posterior mean of $\mu$}  &
\multicolumn{2}{c|}{Posterior mean of $\tau$}  &
\multicolumn{2}{c|}{Bayes factor}                                                                                 \\ \hline
\begin{tabular}[c]{@{}c@{}}$LB_{PM}$\\ $-0.0480$\end{tabular}  & \begin{tabular}[c]{@{}c@{}}$UB_{PM}$\\ $-0.0480$\end{tabular}  & \begin{tabular}[c]{@{}c@{}}$LB_{PM}$\\ $1.0195$\end{tabular}  & \begin{tabular}[c]{@{}c@{}}$UB_{PM}$\\ $1.0205$\end{tabular}  & \begin{tabular}[c]{@{}c@{}}$LB_{PM}$\\ $7.9505$\end{tabular}  & \begin{tabular}[c]{@{}c@{}}$UB_{PM}$\\ $7.9664$\end{tabular}  \\ \hline
\begin{tabular}[c]{@{}c@{}}$LB_{PMA}$\\ $-0.0481$\end{tabular} & \begin{tabular}[c]{@{}c@{}}$UB_{PMA}$\\ $-0.0480$\end{tabular} & \begin{tabular}[c]{@{}c@{}}$LB_{PMA}$\\ $1.0192$\end{tabular} & \begin{tabular}[c]{@{}c@{}}$UB_{PMA}$\\ $1.0208$\end{tabular} & \begin{tabular}[c]{@{}c@{}}$LB_{PMA}$\\ $7.9472$\end{tabular} & \begin{tabular}[c]{@{}c@{}}$UB_{PMA}$\\ $7.9697$\end{tabular} \\ \hline
\end{tabular}
\caption{Lower (LB) and upper bounds (UB) on the Bayes factor and posterior means of $\mu$ and $\tau$.}
\label{tab:BayesFactor}
	\end{table}
	
	\subsection{Bayesian Linear Regression}
	\label{subsec:BayesianLinear}
In this section, we apply the proposed method to a Bayesian linear regression model. Let $y = (y_1, y_2, \dots, y_n)$ be an $n$-dimensional vector denoting the continuous response variable, where $n$ is the number of observations. Let $X$ be an $n \times d$ matrix, where $d$ is the number of covariates and let $\boldsymbol{\beta}$ be a $d$-dimensional coefficient vector of regression parameters. Using matrix notation, the linear regression model can be written as $y = X\boldsymbol{\beta} + e$, where $e \sim N(0,\sigma^2 I_n)$. For Bayesian inference, we assume a vague independent Gaussian prior distribution over all of the unknown regression parameters, $\boldsymbol{\beta} \sim N(0, s_0^2 I_d)$. The true posterior distribution can be easily determined, and is given by:
	\begin{equation*}
	\boldsymbol{\beta}|y \sim N\left(\dfrac{1}{\sigma^2}\left(\dfrac{1}{\sigma^2} X'X + \dfrac{1}{s_0^2} I_d\right)^{-1}X'Y, \left(\dfrac{1}{\sigma^2} X'X + \dfrac{1}{s_0^2} I_d\right)^{-1}  \right).
	\end{equation*}
To assess the performance of our method, we use simulation studies with a varying number of covariates and estimate the $q_i$s for various choices of $\alpha$. For each value of $d$, we generate the design matrix $X$ and the regression coefficients $\boldsymbol{\beta}$ from a continuous uniform distribution, $U(-1,1)$. We then proceed to estimate the unknown regression coefficients using different techniques. Under the proposed $D_\alpha$-based approach, the estimated posterior $q(\boldsymbol{\beta})$ is a product of all of the $q_i$s, $q(\boldsymbol{\beta}) = \prod_{i=1}^{d}q_i(\beta_i)$, and the estimated individual regression coefficients are evaluated using the posterior means corresponding to each $q_i$. We compare our approach to a simple Gibbs sampling algorithm for each value of $d$, and estimate the coefficients using the posterior sample mean after suitable burn-in. To account for the variation in the randomly generated datasets and regression coefficients, we replicate each study $rep$ times. Since the true posterior is known, we calculate the mean squared error (MSE) between the estimator $\hat{\beta}$ and the true value $\beta_{\boldsymbol{\mu}}$: $MSE = \dfrac{1}{rep}\dfrac{1}{d}\sum_{k = 1}^{rep}  \sum_{i = 1}^{d} \left(\hat{\beta}^i_{rep} - \beta^i_{\boldsymbol{\mu} rep}\right)^2$.

Table \ref{tab:linear} reports the results. For each choice of $d$, the estimated regression parameters obtained using the proposed method result in a very small MSE. Although the number of iterations and burn-in is quite large for the Gibbs sampler, it still results in a higher MSE than the $D_\alpha$-based approach. Further, to evaluate the efficiency of our method in a high-dimensional setting, we simulated a single dataset with $d=500$ covariates and $n=1000$ observations. The MSE obtained using the proposed method with $\alpha = 0.5$ was $4.9336\times 10^{-7}$, which shows its utility in the high-dimensional setting.
	
	\begin{table}[!t]
		\centering
		\resizebox{\textwidth}{!}{%
			\begin{tabular}{|c|c|c|c|c|}
				\hline
				\multirow{3}{*}{} & $d  = 25$ & $d = 50$ & $d = 100$ & $d = 200$  \\ \cline{2-5}
				& $n = 100$ & $n = 100$ & $n = 500$ & $n = 500$  \\ \cline{2-5}
				& $rep = 100$ & $rep = 100$ & $rep = 50$ & $rep = 25$  \\ \hline
				\begin{tabular}[c]{@{}c@{}}Gibbs sampling\\ (\textit{iter/burn-in})\end{tabular} & \begin{tabular}[c]{@{}c@{}}7.4368e-07\\ (\textit{50000/20000})\end{tabular} & \begin{tabular}[c]{@{}c@{}}2.7044e-06\\ (\textit{50000/20000})\end{tabular} & \begin{tabular}[c]{@{}c@{}}1.0359e-07\\ (\textit{60000/25000})\end{tabular} & \begin{tabular}[c]{@{}c@{}}2.7866e-07\\ (\textit{60000/25000})\end{tabular} \\ \hline
				$\alpha = 0.5$ & 2.8065e-11 & 3.9600e-10 & 9.4248e-12 & 4.3790e-09 \\ \hline
				$\alpha = 0.9$ & 9.1023e-11 & 9.5112e-10 & 3.9182e-11 & 1.8072e-08 \\ \hline
				$\alpha = 1.1$ & 1.6681e-10 & 1.8017e-09 & 9.9131e-11 & 4.5736e-08 \\ \hline
			\end{tabular}%
		}
		\caption{MSE for Gibbs sampler and $D_\alpha$-based VB for $\alpha = 0.5,\ 0.9,\ 1.1$. $d$: number of unknown regression parameters, $n$: sample size, $rep$: number of simulated datasets for each choice of $d$ and $n$, $\sigma^2 = 1$, $s_0^2 = 100^2$.}
		\label{tab:linear}
	\end{table}

The true marginal is also available in closed form for the Bayesian linear regression setup: $y \sim N\left(0, {\sigma^2} I_n + {s_0}^2 X'X \right)$. We use Proposition \ref{prop:lower_bound} to compute bounds on the logarithm of the marginal. For evaluating the high-dimensional integrals in $\mathcal{H}(f,q)$ and $\mathcal{E}_\alpha(\psi_q; \cdot)$ for KLD-based VB and $D_\alpha$-based VB, respectively, we use the proposed first-order Taylor approximation technique. Let $LB_{KLDA}$ denote the lower bound obtained using the KLD-based VB framework, and $LB_{PMA}$ and $UB_{PMA}$ denote the lower and upper bounds obtained using the proposed methodology with $\alpha = 0.9$ and $\alpha = 1.1$, respectively (these bounds are again computed using the method discussed in Section \ref{subsec:Estimation}). Table \ref{tab:marginalBounds} reports the results for different choices of $d$ and $n$. In all cases, the PMA lower bound is tighter than the KLDA lower bound, with highest differences seen when $d$ is large. The upper bound provided by PMA is also close to the true value of the log-marginal. One could potentially use the average of the lower and upper bounds as an estimate of the true value.

	\begin{table}[!t]
		\centering
		\begin{tabular}{|c|c|c|c|c|c|}
			\hline
			$d$  & $n$   & $LB_{KLDA}$ & \begin{tabular}[c]{@{}c@{}}$LB_{PMA}$ \\ $\alpha = 0.9$\end{tabular} & \begin{tabular}[c]{@{}c@{}}$UB_{PMA}$ \\ $\alpha = 1.1$\end{tabular} & True log marginal \\ \hline
			3  & 10  & -27.5491    & -27.5481 & -27.2727 & -27.5285          \\ \hline
			5  & 20  & -54.2927    & -54.2899  & -53.6364  & -54.0821          \\ \hline
			20 & 100 & -273.5986   & -273.5864 & -272.7273 & -273.0204         \\ \hline
			20 & 200 & -425.8311   & -425.7470  & -425.4545 & -425.8824         \\ \hline
			50 & 250 & -695.2683   & -695.2685  & -694.5455 & -694.8856         \\ \hline
		\end{tabular}
		\caption{Lower (LB) and upper bounds (UB) on the logarithm of the marginal using KLD- and $D_\alpha$-based VB.}
		\label{tab:marginalBounds}
	\end{table}
	
	\begin{figure}[!t]
		\centering
		\resizebox{\columnwidth}{!}{
			\begin{tabular}{|c|c|c|}
				\hline
&\\
				\LARGE{$d = 50, n = 100$} & \LARGE{$d = 100, n = 500$}\\ 
				\hline
				\includegraphics[width=\columnwidth]{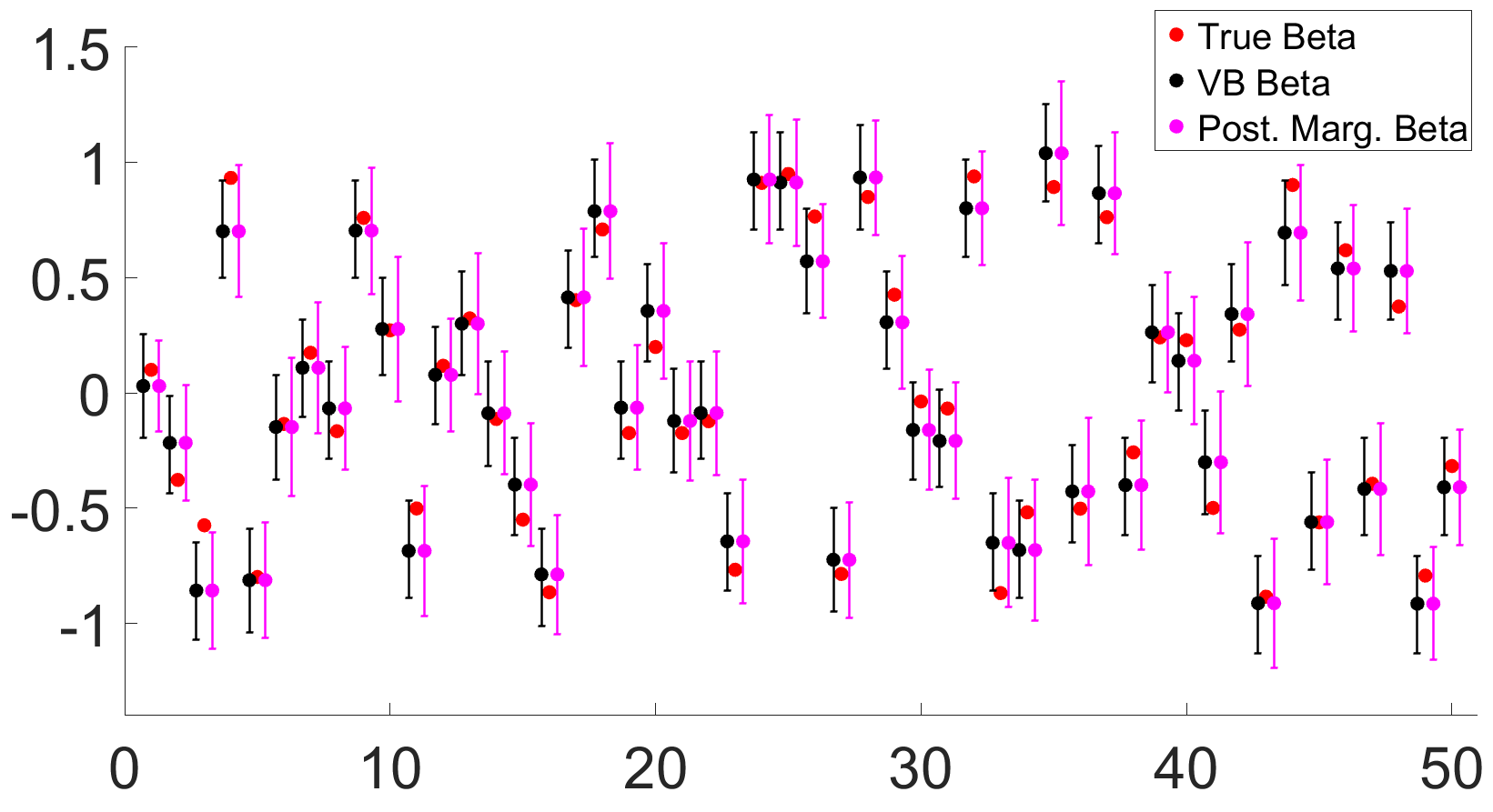} & \includegraphics[width=\columnwidth]{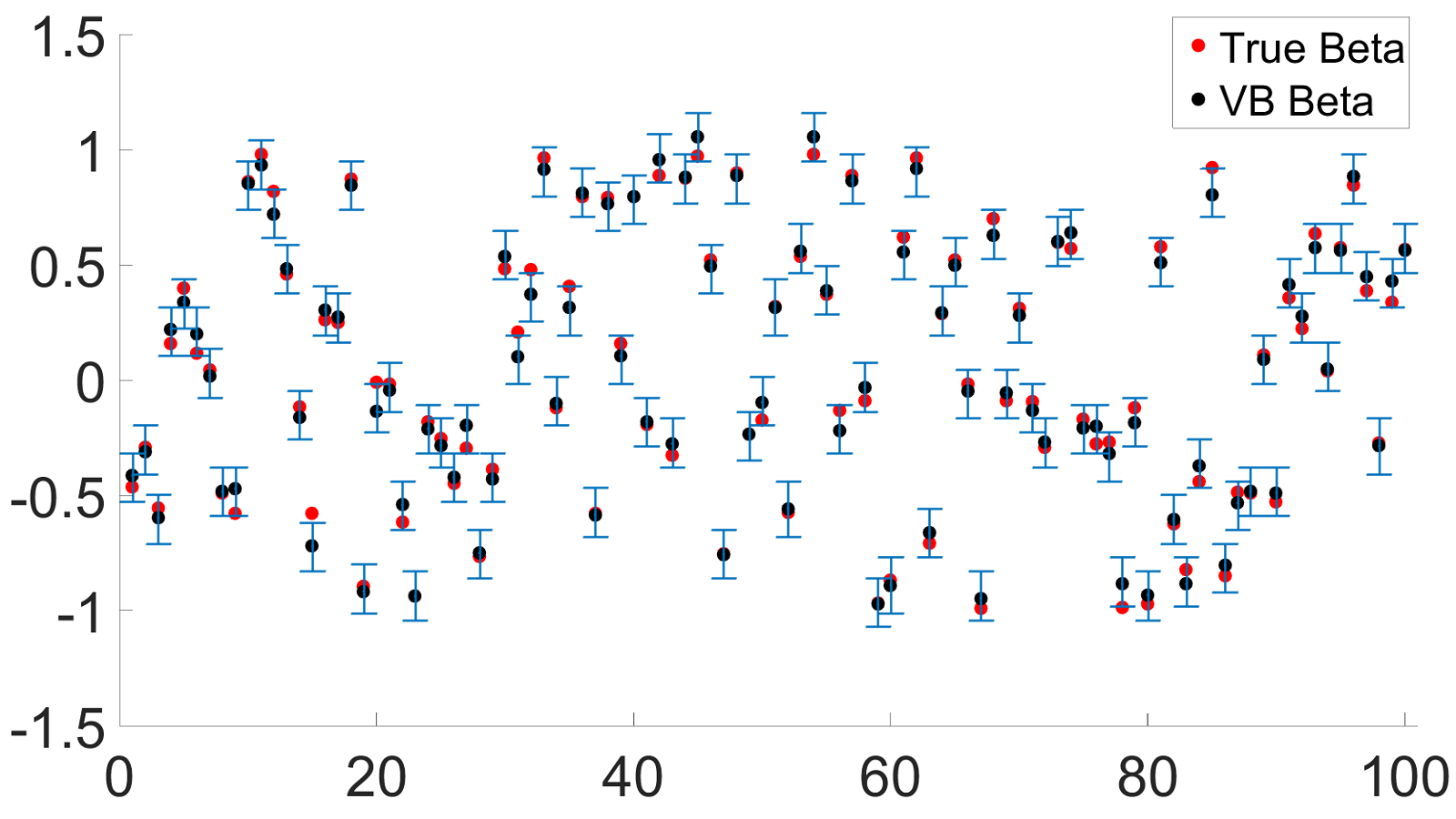}\\
				\hline
			\end{tabular}
		}
		\caption{Equal-tailed $95\%$ posterior credible intervals for $\alpha = 0.5$.}
		\label{fig:postcredintervals}
	\end{figure}
	
Finally, in Figure \ref{fig:postcredintervals}, we report $95\%$ equal-tailed posterior credible intervals based on the estimated $q_i$s using the proposed method with $\alpha=0.5$. For the first example ($d = 50, n = 100$), we also plot the true posterior credible intervals. The $x$-axis represents the regression parameter number and the $y$-axis represents the value of the parameter. In all cases, the intervals calculated using the proposed method do a good job of capturing the value of the true regression coefficient, including the second example ($d = 100, n = 500$), where the intervals get smaller due to a larger sample size. In the left panel of Figure \ref{fig:postcredintervals}, we see that the credible intervals calculated using the proposed method significantly overlap with the true posterior credible intervals. The intervals based on the proposed method are generally shorter in length, albeit not by very much, as compared to the true intervals. This is expected since VB methods tend to underestimate posterior variability \citep{blei2016}. 
	
	\subsection{Bayesian Density Estimation}
	\label{subsec:BayesianDensity}
Logistic Gaussian process (LGP) priors \citep{leonard1978} have been efficiently used as a flexible tool for Bayesian nonparametric density estimation. Theoretical properties of this model have been studied extensively \citep{tokdar2007,van2009}. Further, a quick approximation using Laplace's method for LGP density estimation and regression was proposed in \cite{riihimaki2014}. The resulting posterior distribution obtained using the LGP prior is analytically intractable because of the integral term, which appears in the likelihood function. Before proceeding to show how the proposed method can be used in this setting, we briefly review the LGP model.

Let $x_1, x_2, \dots, x_n$ denote a random sample of size $n$ drawn from an unknown univariate density function, $f$. Let $\mathcal{X}$ denote the support of the distribution. To estimate $f$, we use the logistic density transform \citep{leonard1978} $f(x) = e^ {g(x)}/\int_\mathcal{X} e^ {g(x)} dx$, where $g$ is an unconstrained function. Thus, the problem of estimating the unknown density function $f$ reduces to estimating the function $g$. This transformation is useful as it introduces two necessary constraints for $f$ to be a valid pdf: $f(x) > 0$ and $\int_{\mathcal X} f(x) dx = 1$. To estimate the function $g$, we use a basis expansion model, i.e., $g(x) = \sum_{i = 1}^{d} c_i b_i(x)$, where $c_i$s are the basis coefficients, $b_i$s are the basis functions, and $d$ denotes the number of basis functions used to estimate $g$. We place a noninformative Gaussian prior $\pi_i$ on the unknown coefficients: $c_i \sim N(0, s_0^2),\ \forall\ i = 1,\dots, d$. Let $\boldsymbol{x} = (x_1,\dots, x_n)$ and $\boldsymbol{c} = (c_1,\dots, c_d)$. The joint density function can then be written as $f(\boldsymbol{x},\boldsymbol{c}) = \prod_{j = 1}^{n} f(x_j|\boldsymbol{c}) \prod_{i = 1}^{d} \pi_i(c_i)$, where $f(x_j|\boldsymbol{c}) = \dfrac{exp\{\sum_{i=1}^{d} c_i b_i(x_j) \}} {\int_\mathcal{X} exp\{\sum_{i=1}^{d} c_i b_i(x_j) \} dx}$.

We then use the proposed method to approximate the posterior $p(\boldsymbol{c}|\boldsymbol{x})$ using $q(\boldsymbol{c})$, where $q(\boldsymbol{c}) = \prod_{i = 1}^{d} q_i(c_i)$. Once the approximation to the posterior distribution for each coefficient $c_i$ has been obtained, we calculate the posterior mean, $\hat{c_i},\ \forall\ i = 1, \dots, d$. The expression for the estimated density function is finally given by: $\hat{f}(x) = \dfrac{exp\{\sum_{i=1}^{d} \hat{c_i} b_i(x) \}} {\int_\mathcal{X} exp\{\sum_{i=1}^{d} \hat{c_i} b_i(x) \} dx}.$

\begin{figure}[!t]
	\centering
	\resizebox{0.95\columnwidth}{!}{
		\begin{tabular}{|c|c|}
			\hline
			& \\
			\LARGE{Mixture of N(-1.5,0.25) and N(1.5,1) with equal weights} & \LARGE{Beta(2,5)} \\
			\includegraphics[width=\columnwidth]{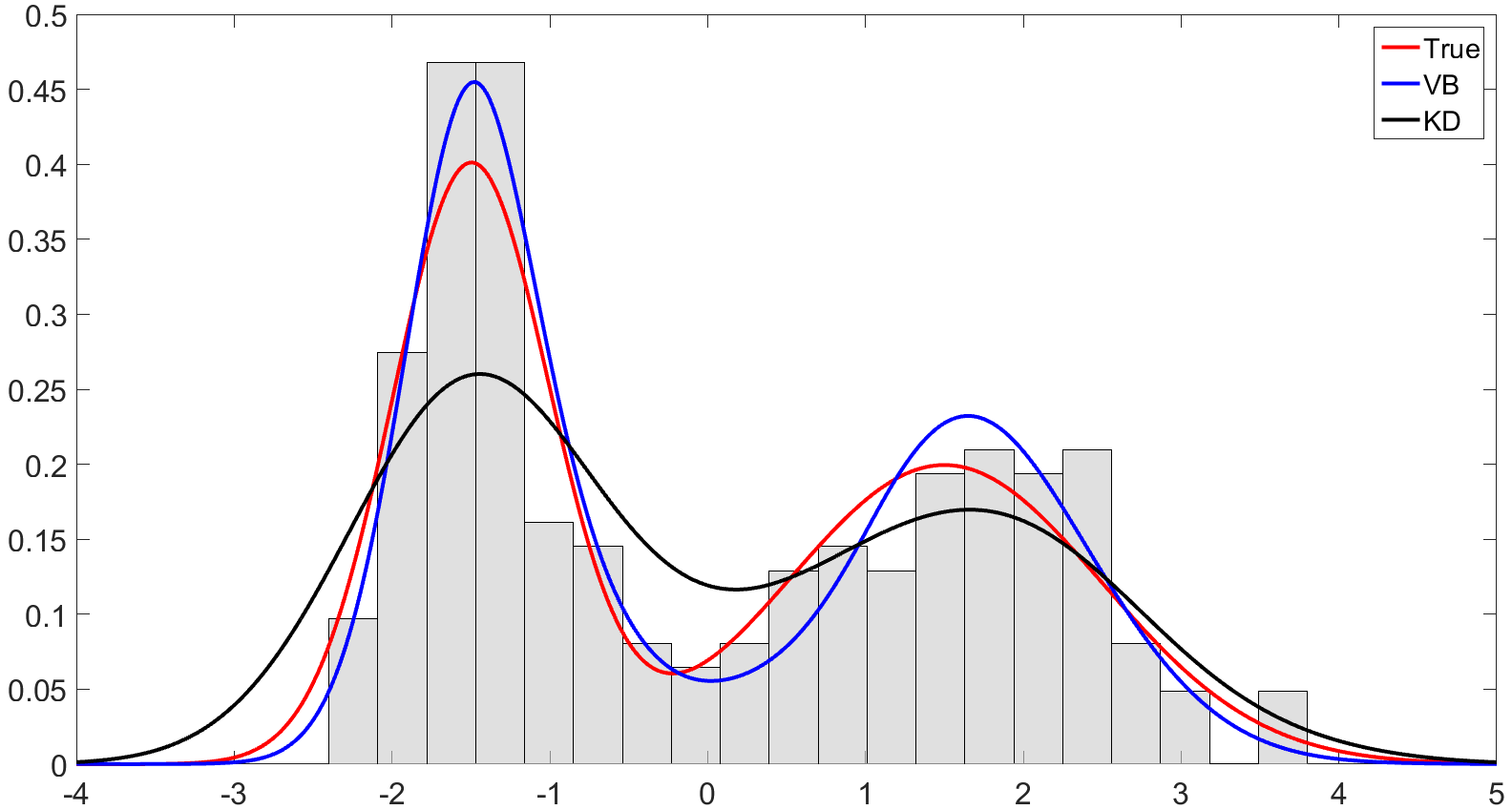} & \includegraphics[width=\columnwidth]{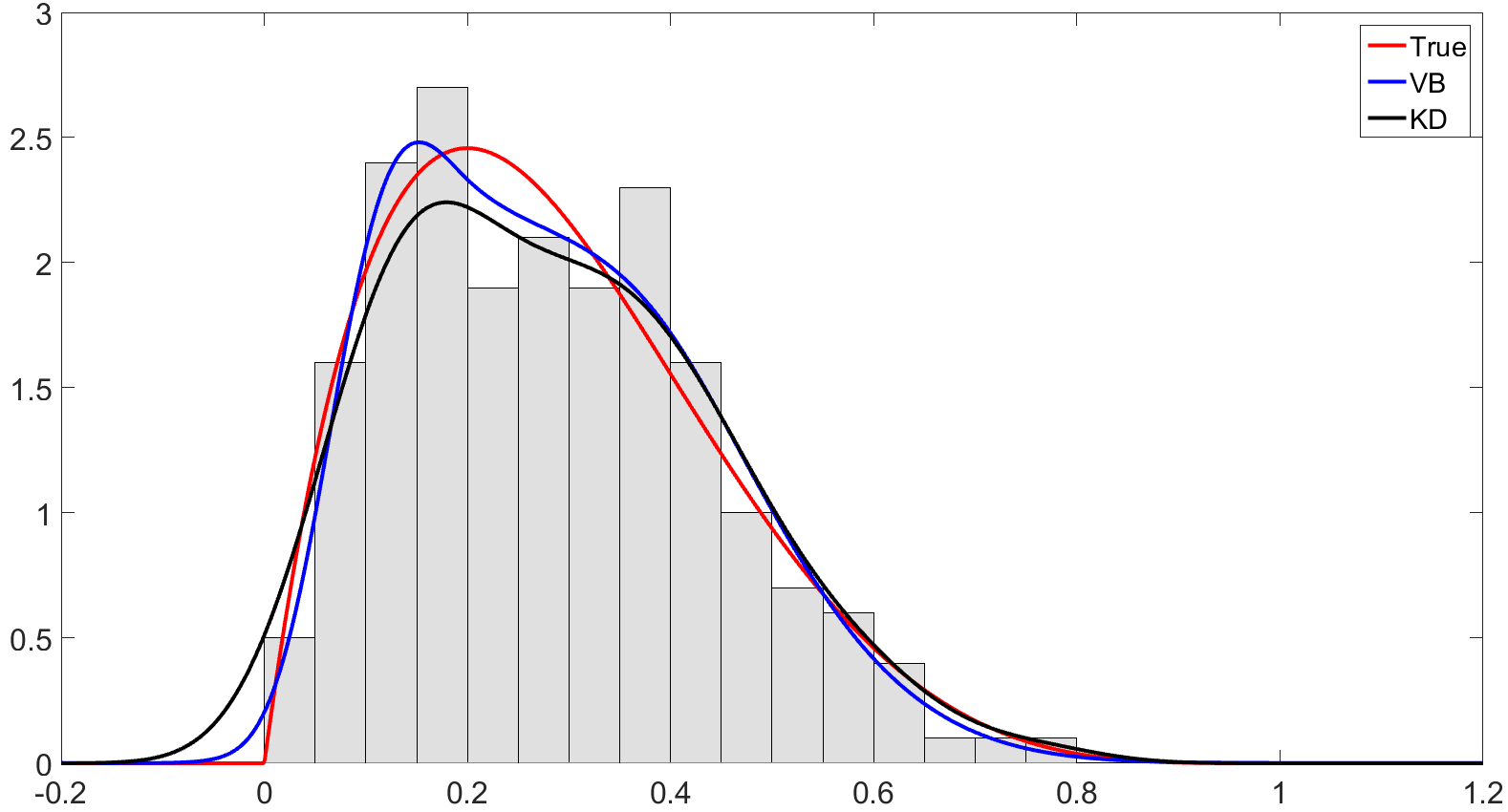} \\ \hline
			& \\
			\LARGE{Uniform(-3,3)} & \LARGE{Beta(0.5,0.5)} \\
			\includegraphics[width=\columnwidth]{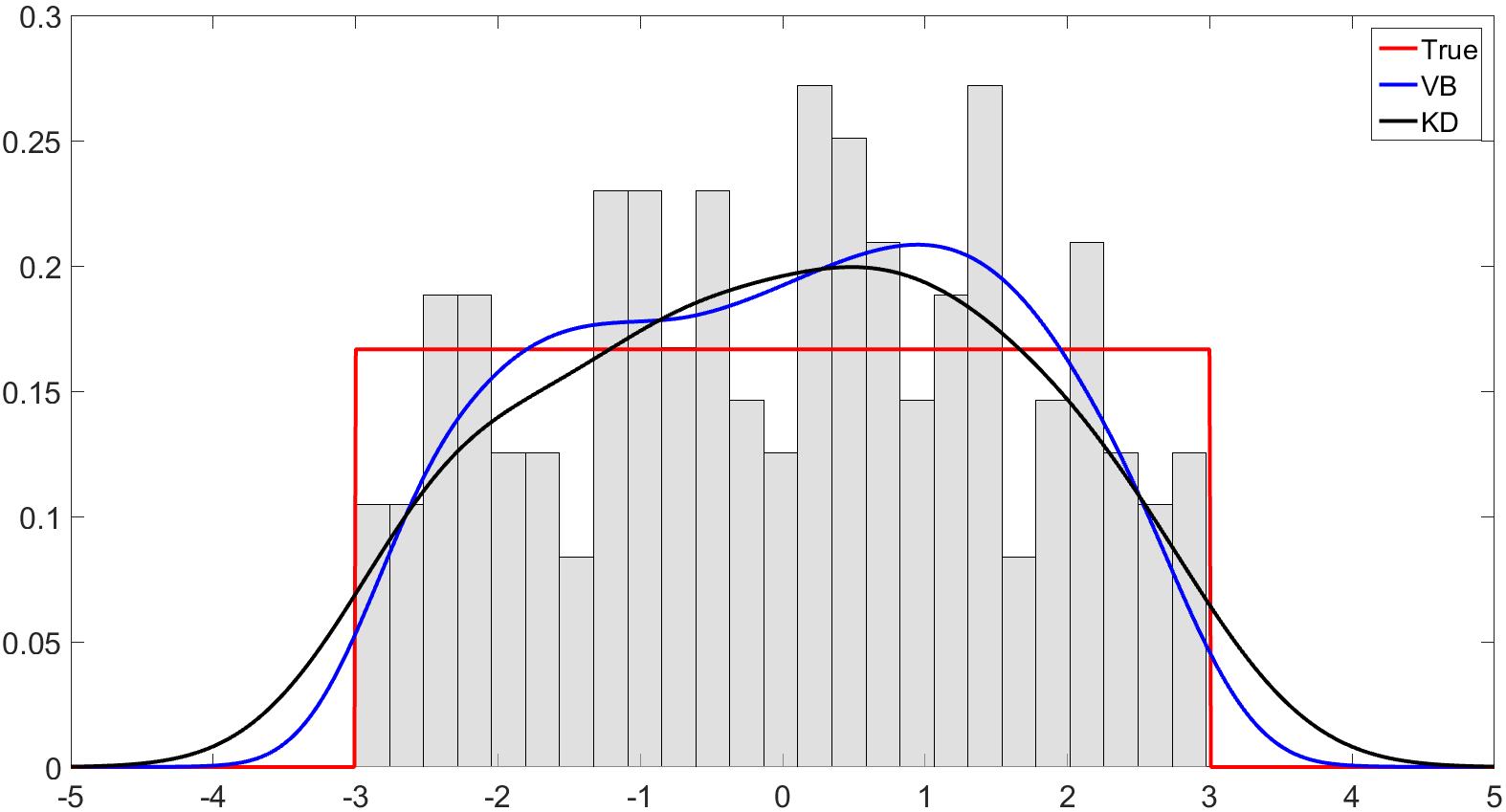} &
			\includegraphics[width=\columnwidth]{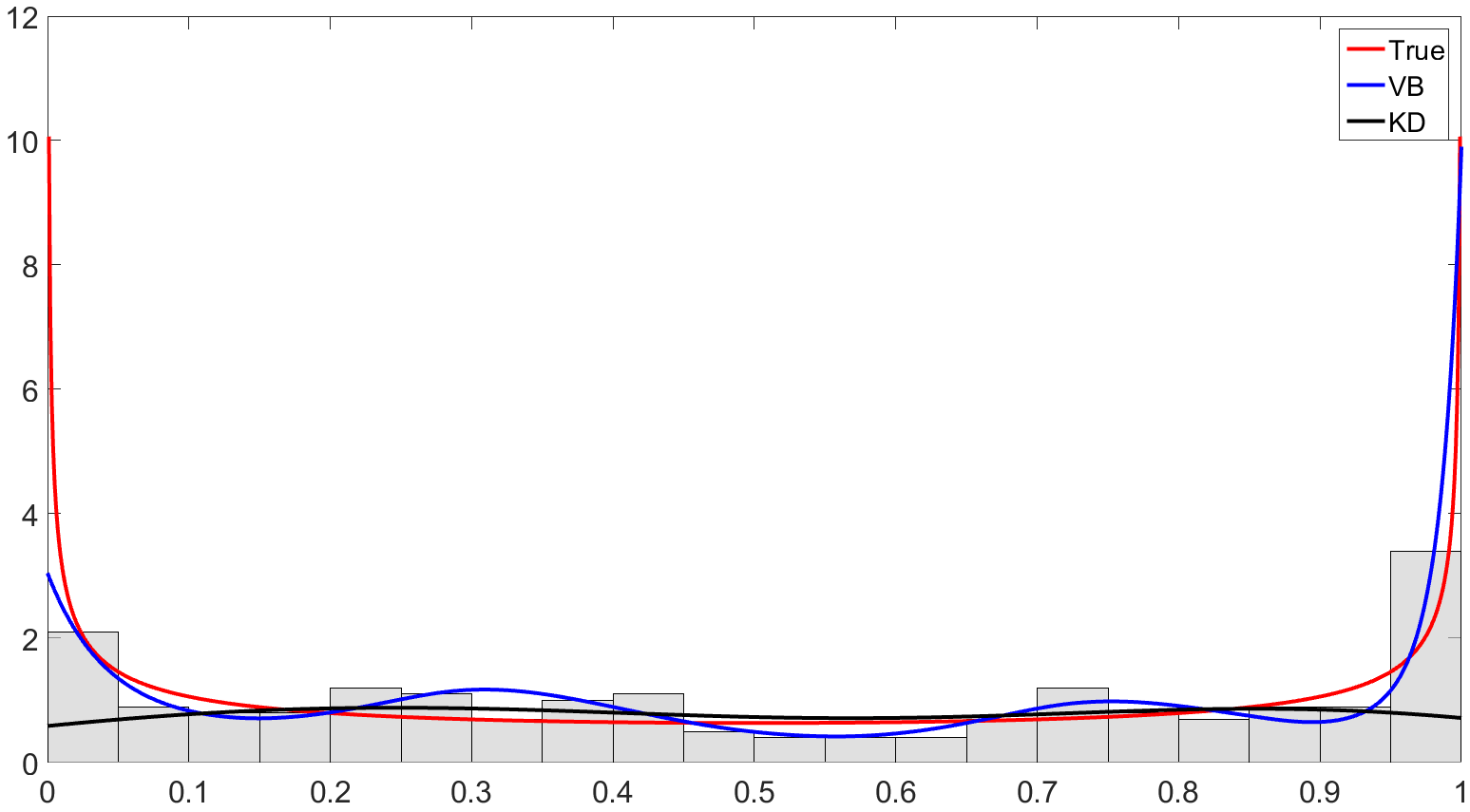} \\ \hline
			& \\
			\LARGE{Gamma(1,2) or Exponential(2)} & \LARGE{Beta(5,1)} \\
			\includegraphics[width=\columnwidth]{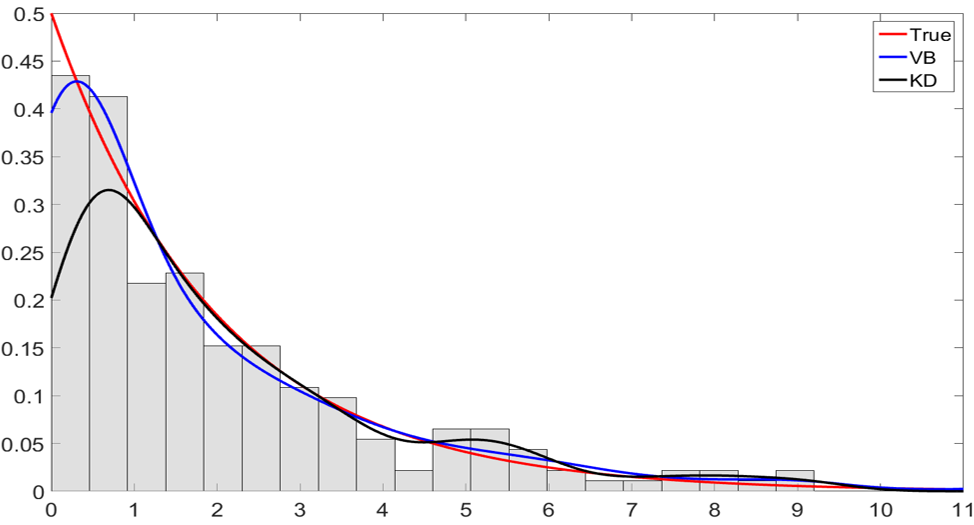} & \includegraphics[width=\columnwidth]{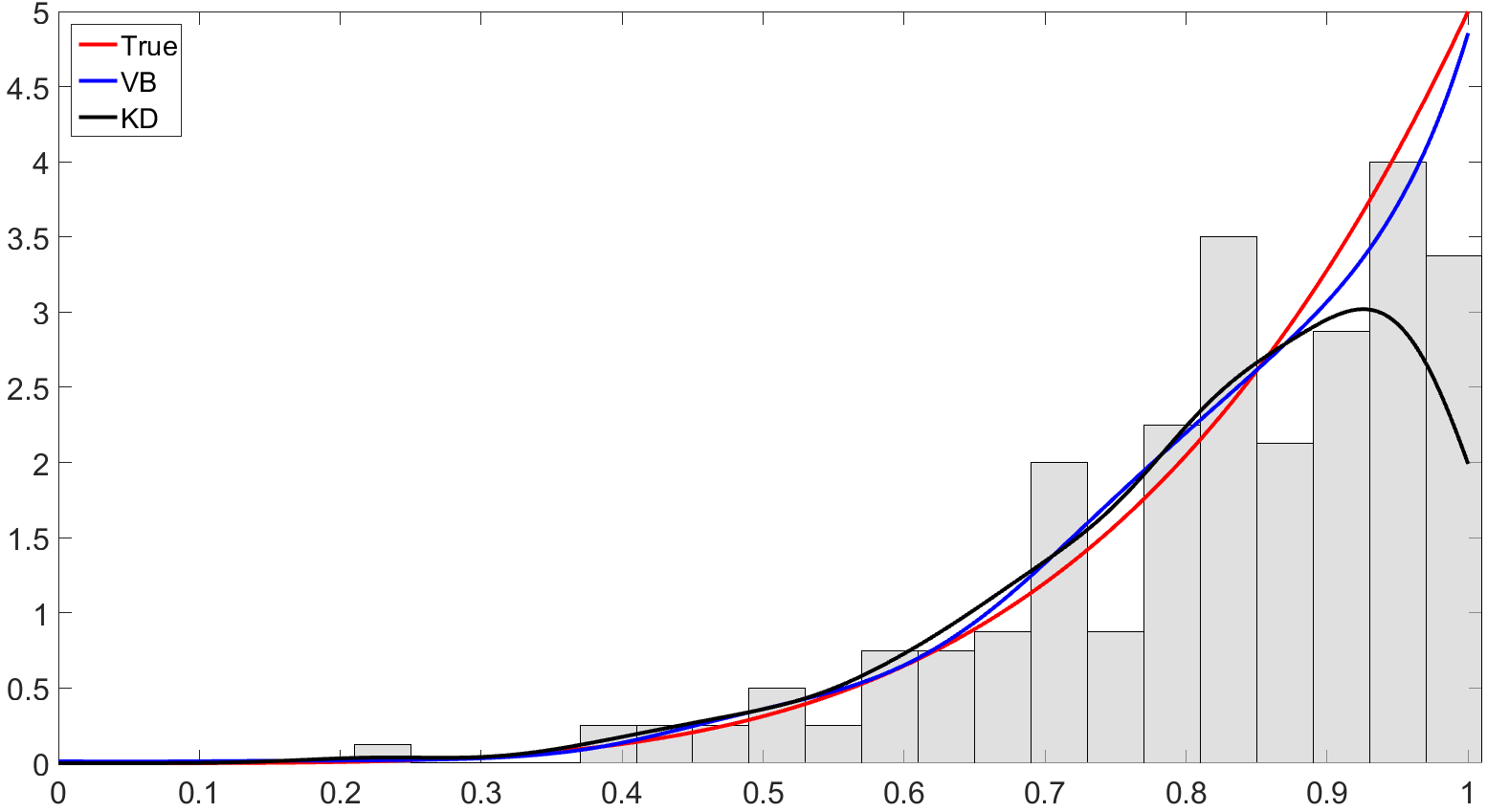} \\ \hline
            & \\				
\LARGE{Gamma(1,2) or Exponential(2)} & \LARGE{Beta(5,1)} \\
				\includegraphics[width=\columnwidth]{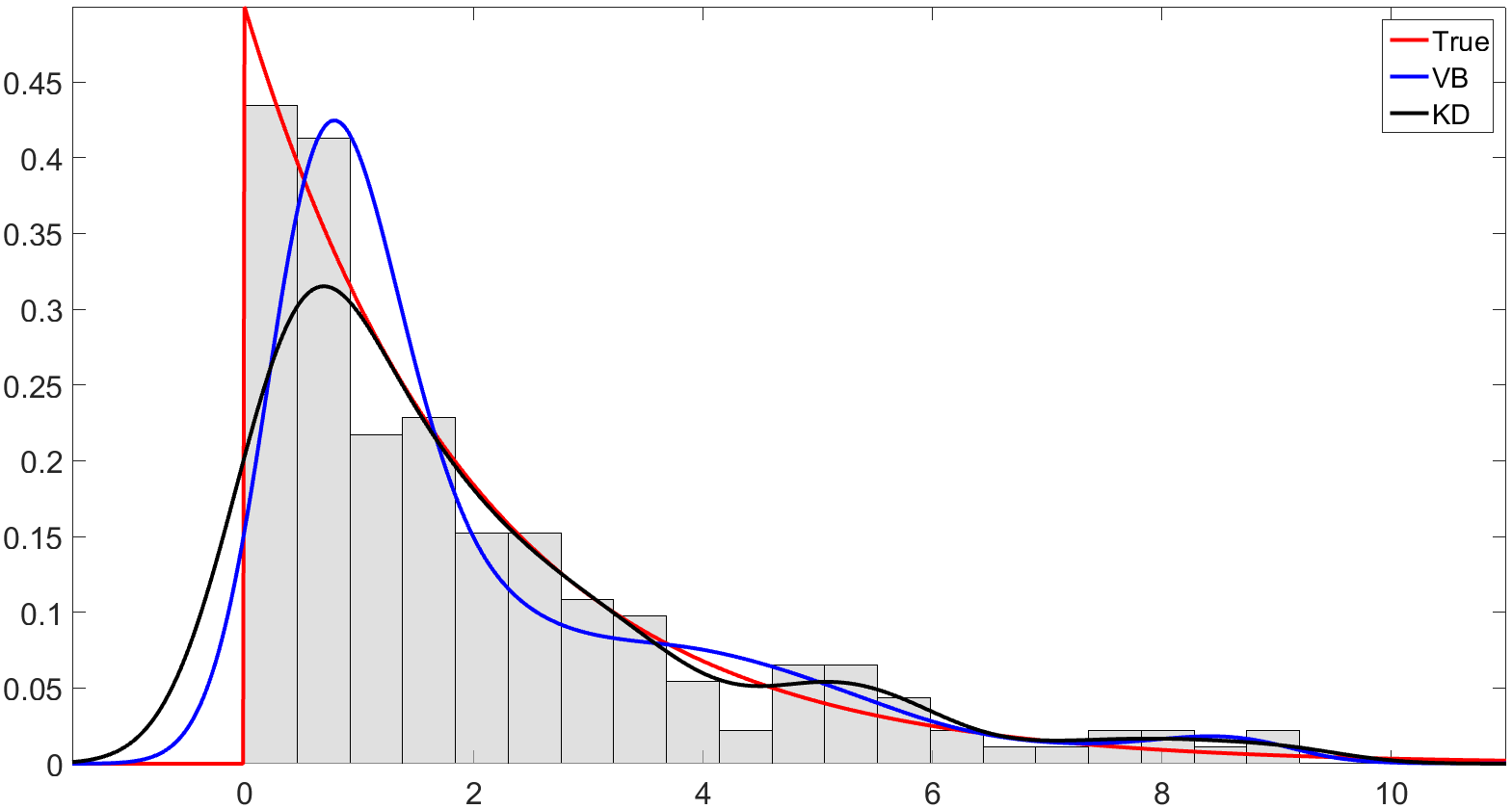} & \includegraphics[width=\columnwidth]{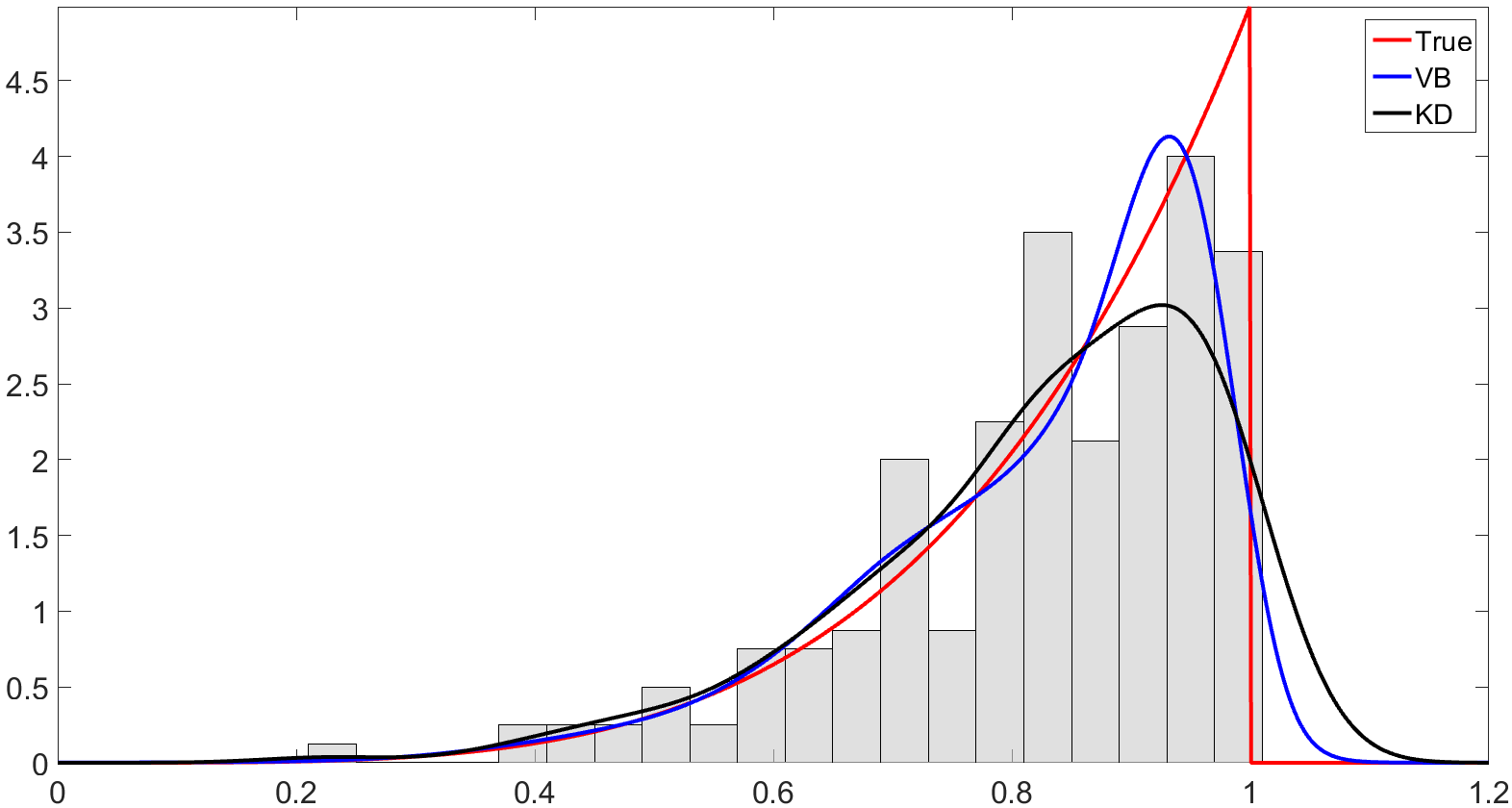} \\
				\hline
		\end{tabular}
	}
	\caption{Bayesian density estimation for various density functions. The simulated data is displayed as a histogram with a plot of the true density (red), density estimate using the proposed method with $\alpha=0.5$ (blue), and a kernel density estimate (black).}
	\label{fig:density}
\end{figure}

\begin{figure}[!t]
		\centering
		\resizebox{\columnwidth}{!}{
			\begin{tabular}{|c|c|c|}
				\hline
				 & & \\
				\LARGE{$d = 5$} & \LARGE{$d = 25$} & \LARGE{$d = 50$} \\
				\includegraphics[width=\columnwidth]{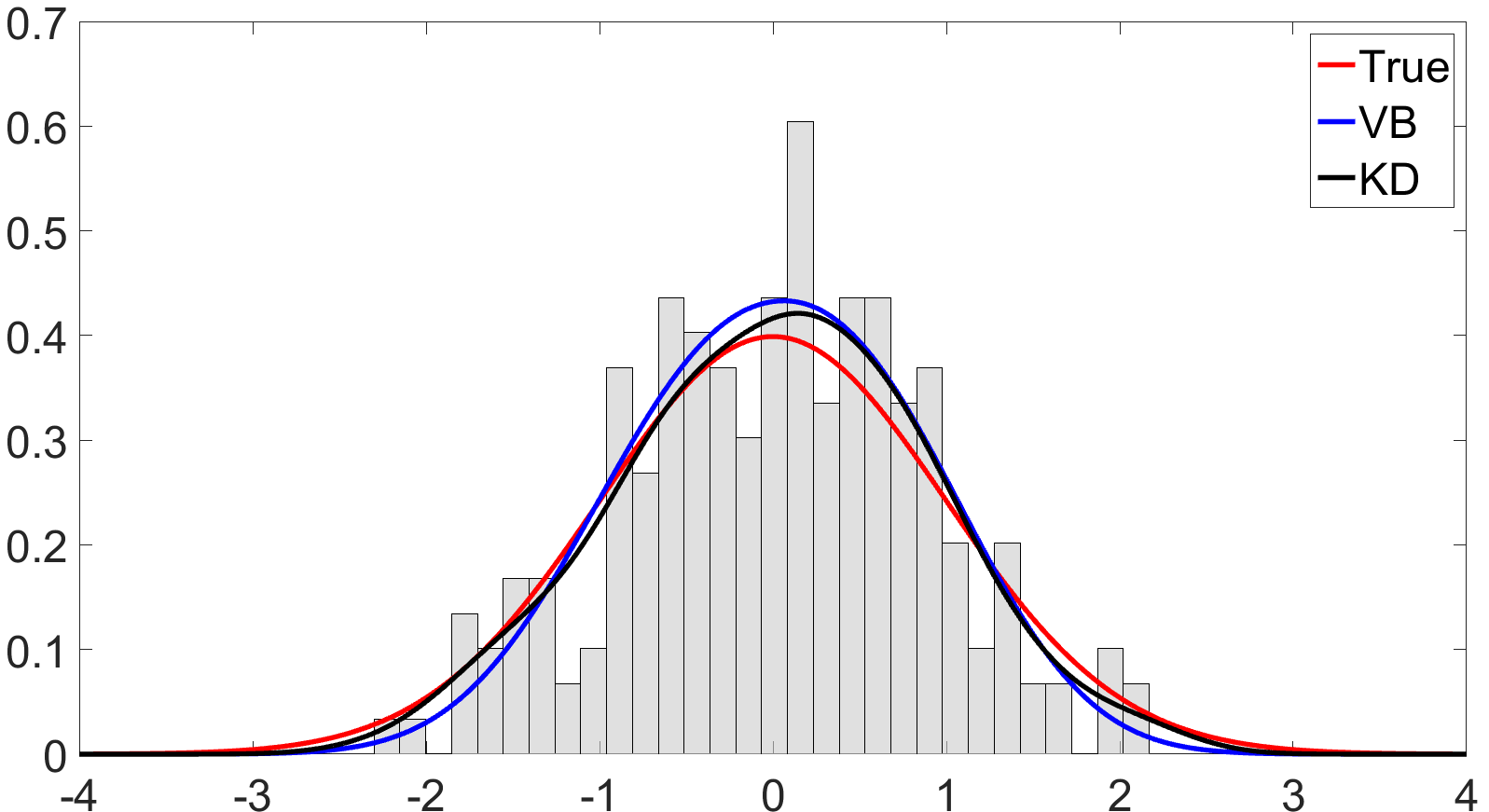} & \includegraphics[width=\columnwidth]{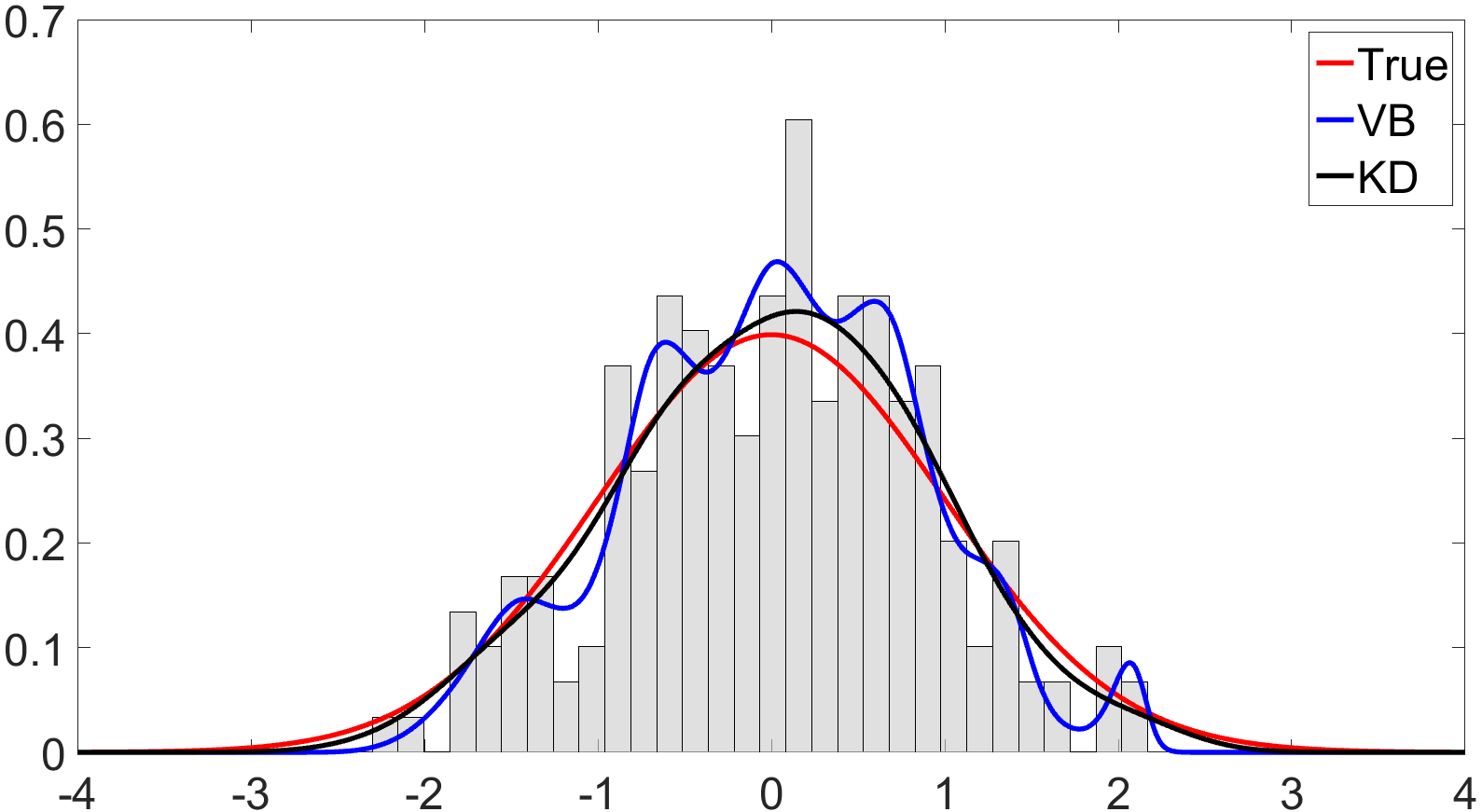} &
				\includegraphics[width=\columnwidth]{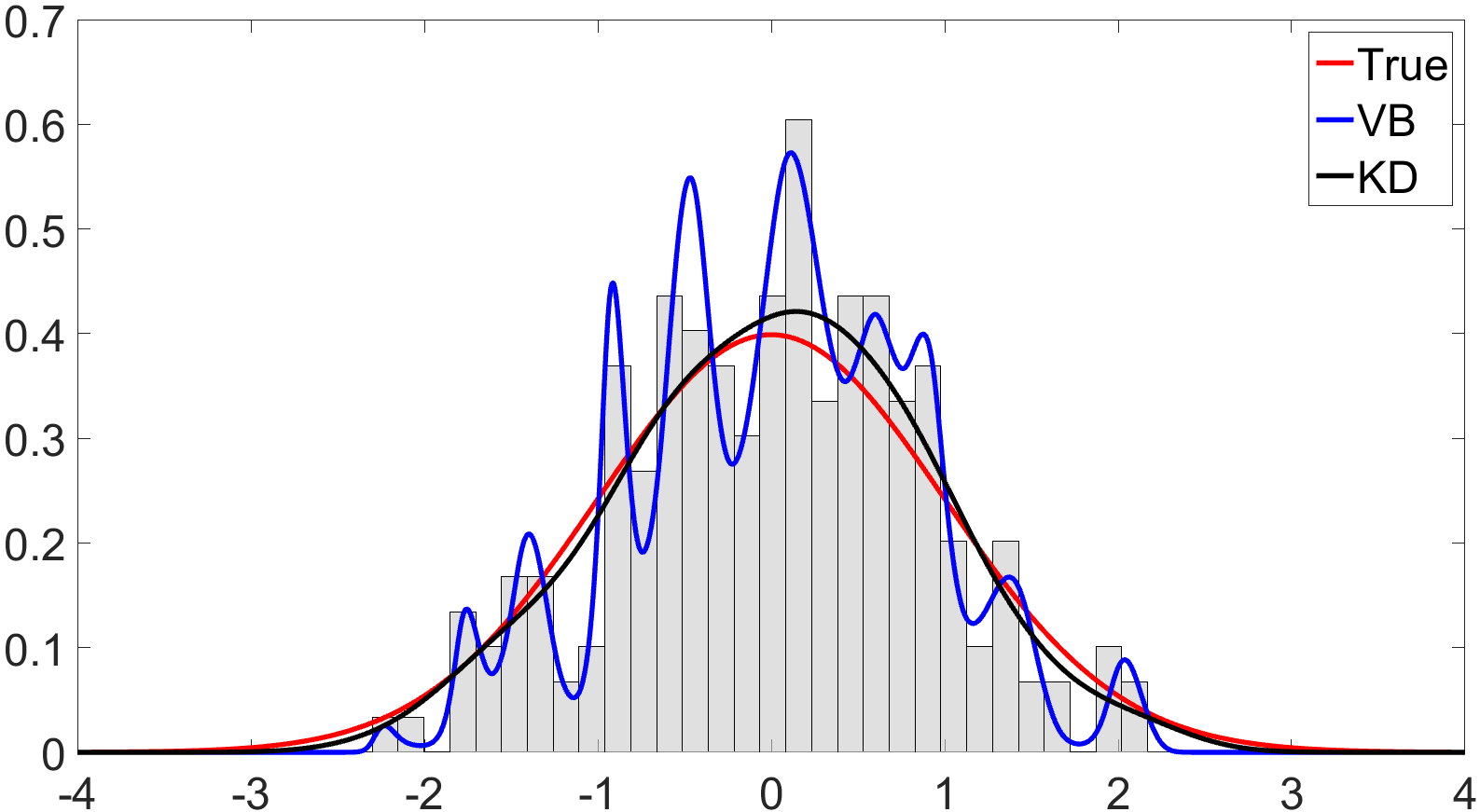} \\
				\hline
			\end{tabular}
		}
		\caption{Effect of increasing the number of basis elements on density estimation. Data was generated from a N(0,1).}
		\label{fig:effect of number of basis elements}
	\end{figure}

To validate the efficiency of our method for estimating density functions, we performed several simulation studies. A random sample was generated from the true underlying distribution in each case, and a histogram corresponding to the sample was used to represent the data. In all of the figures in this section, we plot the true density function in red and the estimated density function in blue. Based on the random sample, we also plot the kernel density estimate in black to provide a visual comparison between the two estimation techniques. The value of $\alpha$ for the proposed method in Figures \ref{fig:density} and \ref{fig:effect of number of basis elements} was chosen to be 0.5. Among the multiple choices for basis functions that can be used to estimate the function $g$, we used B-splines of order four for all of the simulation studies; we also found that Fourier basis provided comparable results. A set of MATLAB code files, supplemental to the book by \cite{ramsay2009} is available for download, and was used to generate the basis functions for all of the examples.

First, we generated datasets from various distributions which exhibit different features as shown in Figure \ref{fig:density}. The third row of Figure \ref{fig:density} shows two plots generated from a Gamma distribution and Beta distribution in the left and right panels, respectively. While implementing our algorithm for density estimation in this example, we set the lower bound for density estimation as 0, since the support of the Gamma distribution is $(0, \infty)$. Similarly, owing to the support of the Beta distribution, i.e., $[0,1]$, we set the lower and upper bounds as 0 and 1, respectively. In practice, the support of the distribution may be unknown; thus, in the last panel we show the same results but without the use of information about the support of the true density. In all cases, the proposed method performs very well compared to standard kernel density estimation.


Figure \ref{fig:effect of number of basis elements} shows the effect of increasing the number of B-spline basis functions used to model $g$. The number of basis functions, $d$, used for estimating the density has a large impact on the final estimate, and behaves similarly to the bandwidth parameter in the kernel density estimator. As we increase the value of $d$, the smoothness of the resulting estimate decreases, and we tend to overfit the data.

%
\subsection{Bayesian Logistic Regression for Real Data Applications}
\label{applications}
We examine the performance on the proposed methodology on binary classification problems using Bayesian logistic regression models. Our choice is motivated by the fact that this is a nonconjugate model that does not fit into the VB setup with conjugate updates. \cite{jaakkola1997} considered variational methods for such models and extended them to binary belief networks. We illustrate that the performance of the proposed geometry-based method is comparable to other approximations, and even better in certain scenarios.

First, we give a brief description of the problem and our classification scheme based on the $D_\alpha$ framework. Let $X$ be a $d \times n$ matrix, where $d$ is the number of covariates (features) and $n$ is the number of observations (cases). Also, let $\boldsymbol{\theta}$ be a $d$-dimensional coefficient vector and $y$ be an $n$-dimensional vector of class labels corresponding to the observations. The class labels take binary values in $\{-1, 1\}$. Under this setup, the logistic regression model is given by $P(y|X,\boldsymbol{\theta}) = g(\boldsymbol{\theta}^{T}X)$ for class label $y = 1$, and $P(y|X,\boldsymbol{\theta}) = g(-\boldsymbol{\theta}^{T}X)$ for class label $y = -1$, where $g(r) = \dfrac{exp (r)}{1 + exp (r)}$. Our final goal is to estimate $\boldsymbol{\theta}$, the vector of unknown coefficients. We again assume vague independent Gaussian priors over all of the unknown parameters in this setting, in the same manner as in Section \ref{subsec:BayesianLinear}. Since the posterior under this setup does not have a closed form expression, we approximate it using $q(\boldsymbol{\theta}) = \prod_{i=1}^d q_i(\theta_i)$ via the proposed variational approach. Finally, for classification purposes, we need to compute the probability $P(y|X,\boldsymbol{\theta})$. There exist various choices based on different features of the posterior that can be used in this scenario; we calculate the following summaries: maximum a posteriori (MAP), posterior mean (PMEA), posterior median (PMED) and posterior predictive (PPRED). If the optimality criterion is chosen to be KLD instead of $D_\alpha$, we can still use the proposed gradient-based algorithm to approximate the posterior. Thus, all of the aforementioned summaries (KLMAP, KLPMEA, KLPMED and KLPPRED) can be obtained using the proposed algorithm for a standard KLD VB framework as well. We use this approach for comparison to $D_\alpha$ and present classification results in terms of accuracy (in $\%$) for each of the methods.

For both of the examples that follow, we use a training set to approximate the posterior distribution of the coefficient vector. We then separately use the four summaries mentioned above to predict the binary class label in a test dataset, and evaluate the classification accuracy. We select a threshold for the binary partition, which minimizes the training error rate based on the posterior predictive in the training set. If the predicted probability is greater than the cutoff, we set $y = 1$, and $y = -1$ otherwise. Further, we also calculate the average log predictive likelihood (ALPL) based on the test set. Given an observation from the test set, we calculate $\log P(y|X,\boldsymbol{\hat{\theta}})$ based on the value of the binary class label $y$, where $\boldsymbol{\hat{\theta}}$ is one of the posterior summaries considered above. A high value of the likelihood signifies better fit of the model.

In the first example, we use a standard benchmark dataset to compare the classification results obtained using the proposed methodology to many other approaches. Further, we compute bounds on the marginal density of the data using the proposed method. In the second example, we apply our approach to the problem of signature verification. We first define a novel set of shape-based descriptors, and then use them as features in a binary genuine vs. forgery classification problem.

    \subsubsection{Ionosphere Data}
    \label{subsub:iono}
The ionosphere dataset \citep{sigillito1989} is a standard binary classification benchmark, which we obtained from the UCI Machine Learning Repository \citep{dua2017}. This data contains 34 predictors corresponding to pulse numbers of signals received by a radar. We remove the second predictor as it is zero for all cases. The binary class labels correspond to good ($y = 1$) or bad ($y = -1$) radar returns. Good radar returns were defined as those showing some type of structure in the ionosphere. There is a total of 351 observations and no missing values.

\begin{table}[!t]
	\centering
	\resizebox{\columnwidth}{!}{
		\begin{tabular}{c|cccc|cccc}
			\hlineB{3}
			& MAP  & PMEA & PMED & PPRED & KLMAP  & KLPMEA & KLPMED & KLPPRED  \\ \hlineB{1}
			Accuracy (in \%) & 96.0 & 96.0 & 96.0 &96.0 & 94.04 & 94.70 & 94.70 & 94.70 \\
			ALPL & -0.1980 & -0.1879 & -0.1979 & -0.1883 & -0.2217 & -0.1886 & -0.2042 & -0.1895 \\
			\hlineB{3}
		\end{tabular}
	}
	\caption{Classification results for the ionosphere dataset.}
	\label{tab:iono}
\end{table}

For classification, we split the full dataset into 200 training and 151 testing cases. We use the same split as reported at \url{http://www.is.umk.pl/~duch/projects/projects/datasets.html#Ionosphere}. This split is very unbalanced: in the training set, the sizes of the two classes are $101\ (50.5\%)$ and $99\ (49.5\%)$, whereas in the test set, the sizes are $124\ (82\%)$ and $27\ (18\%)$, respectively. This website also provides classification results on the same training-testing split for various classification methods. We used the four summaries listed above, both for $D_\alpha$ (with $\alpha = 0.9$) and KLD-based VB, to compute the classification rate. In both cases, we used $499$ basis elements to approximate the energy gradient. Table \ref{tab:iono} presents the results. The proposed method clearly outperforms the KLD-based VB approach, both in terms of classification accuracy and ALPL. We can also compare our results to those listed on the previously mentioned webpage. With six misclassifications, the proposed method ranks fifth best in a list of 23 total methods.

The marginal distribution for the Bayesian logistic regression setup is unavailable in closed form. However, using the same technique as discussed in Section \ref{subsec:BayesianLinear}, we can find bounds on the logarithm of the marginal. For calculating bounds using $D_\alpha$-based VB, we choose $\alpha = 0.9$ and $\alpha = 1.1$ for lower and upper bounds, respectively. The lower bound obtained using KLD-based VB is $-459.5$. Using the proposed method, the lower bound is $-456.7$, and the upper bound is $-448.2$.

    \subsubsection{Application to Signature Verification}
    \label{sub:Signature}

\begin{figure}[!t]
		\begin{center}
			\begin{tabular}{|cc|cc|cc|}
\hline
(a)&(b)&(a)&(b)&(a)&(b)\\
\hline
\includegraphics[width=0.8in]{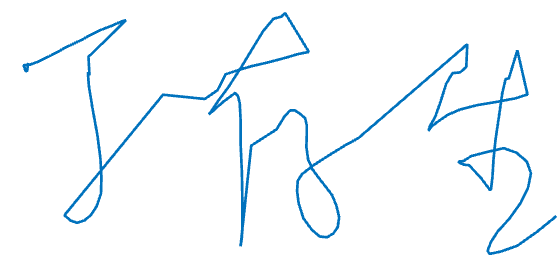}&\includegraphics[width=0.8in]{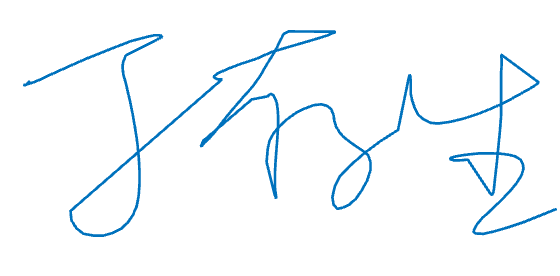}&\includegraphics[width=0.8in]{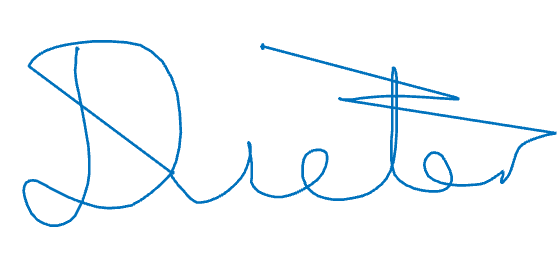}&\includegraphics[width=0.8in]{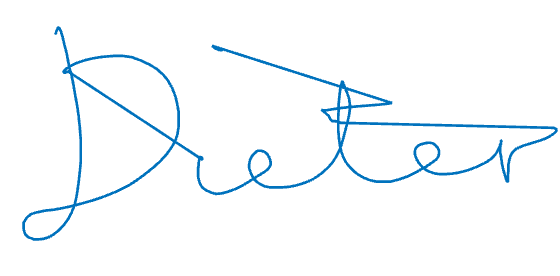}&\includegraphics[width=0.8in]{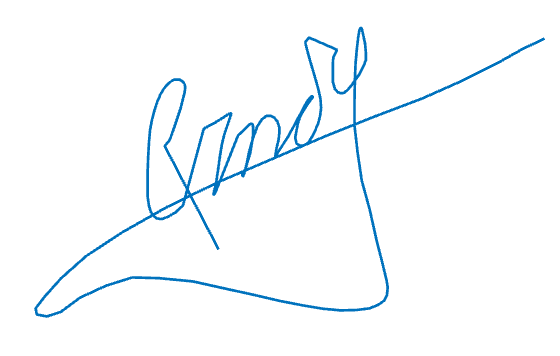}&\includegraphics[width=0.8in]{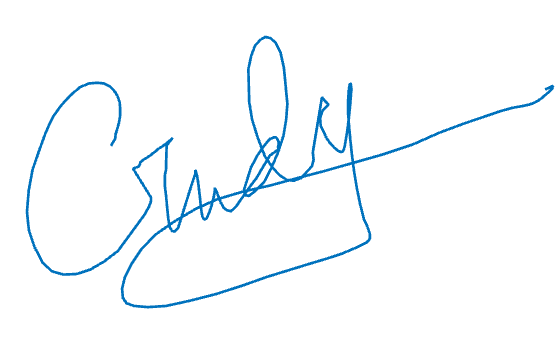}\\
\hline
\end{tabular}
		\end{center}
		\caption{Three examples of (a) genuine and (b) forged signatures.}
		\label{fig:sigs}
	\end{figure}

In this section, we consider the problem of signature verification. The data used here are a subset of the SVC 2004 signature dataset \citep{yeung2004}, which consists of 40 different signatures, each represented by a planar, open curve. For each signature, 20 genuine writing samples and 20 skilled forgeries are provided. We randomly split the data into half training and half testing. We propose to use novel shape-based signature descriptors in conjunction with the proposed variational Bayes framework for this binary classification problem. Figure \ref{fig:sigs} displays four examples of pairs of genuine and forged signatures. The forgeries are extremely difficult to differentiate from the genuine samples making this a difficult classification problem.

To form our descriptors for classification, we use the elastic shape analysis method of \cite{srivastava2011}, which provides tools for registering, comparing and averaging shapes of curves. Let $\beta: [0,1] \rightarrow \mathbb{R}^2$ denote a planar, open, parameterized signature curve. In order to analyze its shape, $\beta$ is represented by a special function, called the square-root velocity function (SRVF) $q:[0,1] \rightarrow \mathbb{R}^2$, defined as $q(t) = \dot{\beta}(t)|\dot{\beta}(t)|^{-1/2}$, where $\dot{\beta} = \frac{d}{dt} \beta$ and $|\cdot|$ is the standard Euclidean norm in $\mathbb{R}^2$. Because the SRVF is defined using the derivative of $\beta$, it is automatically invariant to translation; conversely, $\beta$ can be reconstructed from $q$ up to a translation. In order to achieve invariance to scale, each signature curve is re-scaled to unit length. Because \textit{shape} is a quantity that is invariant to rotation and reparameterization, in addition to translation and scale, these variabilities must also be removed from the representation space. This is performed algebraically using equivalence classes. Let $SO(2)$ be the group of 2 $\times$ 2 rotation matrices (special orthogonal group) and $\Gamma$ be the group of all reparameterizations (orientation preserving diffeomorphisms of $[0,1]$). For a curve $\beta$, a rotation $O \in SO(2)$ and a reparameterization $\gamma \in \Gamma$, the transformed curve is given by $O(\beta \circ \gamma)$. The SRVF of the transformed curve is given by $O(q \circ \gamma) \sqrt{\dot{\gamma}}$. Using this, one can define equivalence classes of the type $[q] = \{ O(q \circ \gamma) \sqrt{\dot{\gamma}}\mid O \in SO(2), \gamma \in \Gamma \}$. Each such equivalence class $[q]$ is associated with a unique shape and vice-versa. Consider two signature curves $\beta_1$ and $\beta_2$, represented by their SRVFs $q_1$ and $q_2$. In order to compare their equivalence classes $[q_1]$ and $[q_2]$, fix $q_1$ and find the optimal rotation and reparameterization of $q_2$ by solving
	\begin{equation}
		\label{eq:sig}
        (O^*, \gamma^*) = \argmin_{O \in SO(2),\ \gamma \in \Gamma} \| q_1 - O(q_2 \circ \gamma) \sqrt{\dot{\gamma}} \|^2.
	\end{equation}
This procedure optimally registers these two shapes. Minimization over the rotation group is performed using Procrustes analysis. Optimization over the reparameterization group requires the dynamic programming algorithm. One can also compute an average shape in this framework using the Karcher mean (minimizer of the sum of squared distances).

To form the signature shape descriptors, we begin by separately computing the average shapes for the genuine and forgery training sets. Next, we register each of the signatures in the training and test sets to both the genuine training average shape and the forgery training average shape using Equation \ref{eq:sig}. For each signature, this results in two different curves $\beta_{gen}^* = O_{gen}^*(\beta \circ \gamma_{gen}^*)$ and $\beta_{for}^* = O_{for}^*(\beta \circ \gamma_{for}^*)$. We then compute the speed functions (magnitude of tangential velocity) defined as $S_{gen}(t) = |\dot{\beta}_{gen}^*(t)|$ and $S_{for}(t) = |\dot{\beta}_{for}^*(t)|$ for each of these curves and concatenate them. The original signature curves are sampled with $100$ points resulting in 200 signature shape descriptors.

For each type of signature, we use the training set to approximate the posterior distribution of the logistic regression model parameters using the proposed variational approach. We use 99 basis elements to approximate the energy gradient with $\alpha = 0.9$. As before, we use summaries of the approximate posterior to compute the classification performance. The results averaged over all test signatures (total of 800) are given in Table \ref{tab:signature}. Note that the proposed shape-based signature descriptors perform extremely well on this signature verification task, both in terms of accuracy and ALPL. Since the split of the training and test set in this case is very balanced, we also present classification results obtained using an empirical cutoff of $0.5$. Interestingly, this choice of cutoff performs better than the results obtained using the minimum training error cutoff based on the posterior predictive. Overall, the proposed method is very successful in this application.

	\begin{table}[!t]
\centering
\begin{tabular}{cc|cccc}
	\hline
	\multicolumn{2}{c|}{}               & MAP         & PMEA    & PMED    & PPRED   \\ \cline{1-2}
	\hline
	\multirow{2}{*}{Accuracy (in \%)}    & (a)   & 100         & 91      & 96.5    & 83.3    \\
	& (b)   & 100         & 99.8    & 99.6    & 99.8    \\
	\hline
	\multicolumn{2}{c|}{ALPL} & -2.8848e-07 & -0.0075 & -0.0115 & -0.0178 \\
	\hline
\end{tabular}
\caption{Classification results for the signature dataset averaged over 40 different signature types for two methods. (a) Minimum training error cutoff and (b) empirical 0.5 cutoff.}
\label{tab:signature}
	\end{table}
	\section{Discussion}
    \label{sec:Summary}

The use of Fisher--Rao Riemannian geometry for the analysis of PDFs has been demonstrated in various settings, including diffeomorphic density matching \cite{DBLP:journals/siamis/BauerJM15}, random sampling via optimal information transport \cite{bauer2017}, sensitivity analysis in Bayesian models \cite{kurtek2015}, and computer vision \cite{srivastava2007}. The unified metric structure and availability of high-speed computing resources provide a natural habitat for the formulation of variational versions of several tasks involving high-dimensional data. Theoretical study of resulting estimates and their comparison with ones currently used within the statistical literature, with a view towards inference, will be highly beneficial.

By moving to the space of nonparametric densities, the availability of explicit expressions for the exponential and inverse-exponential maps under the SRD representation plays a crucial role in the scalability of the proposed gradient ascent algorithm. Our approximation of the gradient direction is based on nested univariate first-order Taylor expansions of a high-dimensional integral; while this worked well in our investigations, better approximation schemes can be explored. There are multiple direction for future work including (1) examination of choice of appropriate basis functions in the tangent space to better capture modalities of the posterior, (2) building upon Proposition \ref{stepsize} to obtain theoretical guarantees for the proposed algorithm (encouragingly, the SRD representation space is a convex subset of the Hilbert sphere, and this will assist us in studying convergence properties), (3) development of efficient initialization schemes for different problems of interest, and (4) extending the proposed framework to a variety of other Bayesian models including generalized linear models, graphical models, spatial models.

\subsection{Extension to Non-mean-field Setting}

We comment now on how the proposed approach can be extended to the setting where we do not assume that joint densities $q$ on $\boldsymbol{\Theta}$ factorize. The definition of the variational family and the square-root map remain unchanged. The definition of the loss function $\mathcal{L}$ can be easily modified to reflect the new variational family. The significant changes lie in the implementation of the proposed algorithm. Recent work by \cite{tan2018arXiv} considered a model-dependent reparameterization trick that can capture posterior dependencies between parameters. The invertible affine transformation that they propose is similar in nature to the reparameterization considered in Proposition \ref{prop:invariance}, and can thus be used in our setting. However, the reparameterization invariance only applies when $\alpha = 1/2$, limiting the applicability of this approach.

The key ingredients of the algorithm are the orthonormal bases, the exponential map, the gradient direction and the parallel transport. The exponential map and the parallel transport can appropriately be modified to reflect the $d$-dimensional nature of the density space. Given the $d$-dimensional basis set of orthonormal basis functions, the expression for the gradient can be written down explicitly. The computation of the gradient however is not straightforward. The key observation here lies in our approximation method based on nested approximations. Denote by $f_{d}(\theta_1):=f(\theta_1|\mu_{2|3,\ldots,d},\mu_{3|4,\ldots,d},\ldots,\mu_d)$ the density $f$ of $\theta_1$ conditioned on the conditional expectations (e.g., $\mu_{2|3,\ldots,d}$ denotes the conditional expectation of $\theta_2$ given $\theta_3,\ldots,\theta_d$), and $\tilde{f}_d(\theta_1):=f(\theta_1,\mu_{2|3,\ldots,d},\mu_{3|4,\ldots,d},\ldots,\mu_d)$.
The resulting modification of Equation \ref{eq:integral} for $i=1$ is $\int_{\Theta_{1}} \left[\dfrac{\psi_f(x, \theta_1, \theta_2, \dots ,\theta_d)}{\psi_{q_d(\theta_1)}} \right]^{2\alpha}\tilde{b}_d^{k}(\theta_1){q}_d(\theta_1) d{\theta_1}$, where $\tilde{b}_d^{k}$ is the $k$th element of the $d$-dimensional orthonormal basis function set. This requires us to compute only one-dimensional conditional expectations. One approach to this is to start with a parametric family for the approximating density and embed it into the nonparametric space of all $d$-dimensional densities. If $d$ is too large, we can consider a more generalized block structure similar to structured mean-field approximation \citep{saul1996,barber1999}. Instead of assuming that all the parameters are mutually independent and controlled by their individual marginals, we can exploit the presence of a substructure in the collection of parameters, assume partial factorization, and continue along the lines mentioned above. The key point however is that the proposed framework can, in principle, be extended to the non-mean-field setting. Much remains to be done in this direction, and is currently work in progress.

\noindent\textbf{Supplementary Material:} The supplementary material includes proofs of all propositions as well as additional results for Bayesian linear regression, Bayesian density estimation and Bayesian logistic regression.

\if0\blind
{
\noindent\textbf{Acknowledgements:} The authors would like to thank Prof. Steven MacEachern for valuable discussions and suggestions. They are also grateful for the comments provided by two anonymous reviewers that improved the contents of this manuscript. This research was partially supported by NSF DMS 1613054 and NIH R01 CA214955-01A1 (to KB and SK), and NSF CCF 1740761 (to SK).
} \fi

\begin{center}
	{\LARGE\bf Supplementary Material for ``A Geometric Variational Approach to Bayesian Inference"}
\end{center}

\section{Proofs}
\noindent\textbf{Proof of Proposition 1.}

Denote by $J (\boldsymbol{\eta}):=\Big|\text{det}\Big(d\theta_i/d\eta_j\Big)\Big|$ the absolute value of the determinant of the Jacobian matrix corresponding to the transformation $(\theta_1,\ldots,\theta_d) \mapsto (\phi_1(\theta_1),\ldots,\phi_d(\theta_d))=(\eta_1,\ldots,\eta_d)$. For $i=1,\ldots,d$ since each $\phi_i$ transforms $\theta_i$ individually, $J(\boldsymbol{\eta})=\prod_{i=1}^d d \theta_i/d\eta_i$. Then,
\begin{align*}
	\mathcal{E}_{1/2} (\psi_q;\boldsymbol{\eta})&= \int_{\boldsymbol{\Theta}} {\psi_f(x,\boldsymbol{\eta})} {\prod_{i = 1}^d \psi_{q_i}(\eta_i)} d\boldsymbol{\eta}
	=\int_{\boldsymbol{\Theta}} {\psi_f(x,\boldsymbol{\theta})}{\prod_{i = 1}^d \psi_{q_i}(\theta_i)} J(\boldsymbol{\eta})d\boldsymbol{\eta}\\
	&=\int_{\boldsymbol{\Theta}} {\psi_f(x,\boldsymbol{\theta})}{\prod_{i = 1}^d \psi_{q_i}(\theta_i)} d\boldsymbol{\theta}=\mathcal{E}_{1/2} (\psi_q;\boldsymbol{\theta}).
\end{align*}
\newline
\noindent\textbf{Proof of Proposition 2.}
\begin{equation*}
	m(x)^\alpha = \Big[ \int_{\boldsymbol{\Theta}} f(x,\boldsymbol{\theta}) d\boldsymbol{\theta} \Big]^{\alpha} = \Big[ \int_{\boldsymbol{\Theta}} \dfrac{f(x,\boldsymbol{\theta})}{q(\boldsymbol{\theta})} q(\boldsymbol{\theta}) d\boldsymbol{\theta} \Big]^{\alpha} = g \Bigg( \mathbb{E}_q \Big[ \dfrac{f(x,\boldsymbol{\theta})}{q(\boldsymbol{\theta})} \Big] \Bigg),
\end{equation*}
where $g(y) = y^\alpha$ is concave on $y>0$ and $0 < \alpha < 1$. Using Jensen's inequality, we obtain:
\begin{equation*}
	m(x)^\alpha \geq  \mathbb{E}_q \Bigg[ g \Big( \dfrac{f(x,\boldsymbol{\theta})}{q(\boldsymbol{\theta})} \Big) \Bigg] = \int_{\boldsymbol{\Theta}} \Big( \dfrac{f(x,\boldsymbol{\theta})}{q(\boldsymbol{\theta})} \Big)^{\alpha} q(\boldsymbol{\theta}) d\boldsymbol{\theta} =  \int_{\boldsymbol{\Theta}} f(x,\boldsymbol{\theta})^{\alpha} q(\boldsymbol{\theta})^{1 - \alpha} d\boldsymbol{\theta},
\end{equation*}
which is equivalent to $\mathcal{E}_\alpha$ since $f = \psi_f^2$ and $q = \psi_q^2$. Taking the natural logarithm on both sides, we obtain $\alpha \ln m(x) \geq \ln \mathcal{E}_\alpha(\psi_q; \cdot)$. Further, consider
$\ln \mathcal{E}_\alpha(\psi_q; \cdot) = \ln \mathbb{E}_q \Bigg[ \Big( \dfrac{f(x,\boldsymbol{\theta})}{q(\boldsymbol{\theta})} \Big)^{\alpha} \Bigg]$. Since $\ln$ is a concave function, using Jensen's inequality again we obtain:
\begin{equation*}
	\ln \mathbb{E}_q \Bigg[ \Big( \dfrac{f(x,\boldsymbol{\theta})}{q(\boldsymbol{\theta})} \Big)^{\alpha} \Bigg] \geq \mathbb{E}_q \Bigg[ \ln \Big( \dfrac{f(x,\boldsymbol{\theta})}{q(\boldsymbol{\theta})} \Big)^{\alpha} \Bigg] = \alpha \mathbb{E}_q \Bigg[ \ln \Big( \dfrac{f(x,\boldsymbol{\theta})}{q(\boldsymbol{\theta})} \Big) \Bigg] = \alpha \mathcal{H}(f,q).
\end{equation*}
Thus, for $0 < \alpha < 1$ we have
\begin{align*}
	\alpha \ln m(x) \geq \ln \mathcal{E}_\alpha(\psi_q; \cdot) \geq \alpha \mathcal{H}(f,q).
\end{align*}
Similarly, for part (ii) of the proposition, noting that $g(y) = y^\alpha$ is convex for $y>0$ and $\alpha > 1$, using Jensen's inequality we obtain:
\begin{align*}
	\alpha \ln m(x) \leq \ln \mathcal{E}_\alpha(\psi_q; \cdot).
\end{align*}
\noindent \textbf{Proof of Proposition 3.}

The directional or Fr\'{e}chet derivative $D^i\mathcal{E}_\alpha$ at a point $\psi_{q_i} \in \Psi$ is defined as a linear functional $T_{\psi_{q_i}}(\Psi)\to \mathbb{R}$, the element of the dual space of $T_{\psi_{q_i}}(\Psi)$, via the relation $D^i\mathcal{E}_\alpha (v_i)=\langle D^i\mathcal{E}_\alpha,v_i\rangle$ for all $v_i \in T_{\psi_{q_i}}(\Psi)$ since $T_{\psi_{q_i}}(\Psi)$ is a linear subspace of $\mathbb{L}^2(\Theta_i)$, the Hilbert space of square-integrable functions on $\Theta_i$, and inherits the usual inner product. The Riesz representation theorem implies that the gradient $\nabla \mathcal{E}_\alpha^i$ exists as an element of $T_{\psi_{q_i}}(\Psi)$ and is defined such that
$D^i\mathcal{E}_\alpha (v_i)=\langle \nabla \mathcal{E}_\alpha^i, v_i \rangle, \quad v_i \in T_{\psi_{q_i}}(\Psi).$
Along a basis direction $b^k_i$, we can therefore express the gradient as
\[
\nabla \mathcal{E}_\alpha^i=\sum_{k=1}^\infty D^i\mathcal{E}_\alpha (b^k_i) b^k_i.
\]
Therefore, on the restriction $\mathcal{E}_{\alpha|\Psi}:\Psi\to \mathbb{R}_{>0}$ to $\Psi$ for each $i=1,\ldots,d$, the directional derivative $D^i\mathcal{E}_\alpha$ along $b^k_i$ can be computed as
\begin{align*}
	&D^i\mathcal{E}_\alpha(b ^k_i)=\lim_{t\to 0}
	\frac{1}{t}\Big[\mathcal{E}_{\alpha|\Psi}(\psi_{q_i}+tb^k_i)-\mathcal{E}_{\alpha|\Psi}(\psi_{q_i})\Big]\\
	&=\lim_{t\to 0}
	\frac{1}{t}\Big[\int_{\boldsymbol{\Theta}}\psi_f(x,\boldsymbol{\theta})^{2\alpha}\prod_{j\neq i}(\psi_{q_j}(\theta_j))^{2(1-\alpha)}\left\{(\psi_{q_i}(\theta_i)+tb^k_i(\theta_i))^{2(1-\alpha)}-(\psi_{q_i}(\theta_i))^{2(1-\alpha)}\right\}d\boldsymbol{\theta}\Big].
\end{align*}
With $p=2(1-\alpha)$, using the binomial expansion for real powers we obtain
\[\frac{1}{t}\Big[(\psi_{q_i}(\theta_i)+tb^k_i(\theta_i))^p\Big]=\frac{1}{t}\Big[\psi^p_{q_i}(\theta_i)+p\psi^{p-1}_{q_i}(\theta_i)tb^k_i(\theta_i)+R(\theta_i,\alpha,t)\Big],
\]
where $R(\theta_i,\alpha,t)=O(m(\theta_i,\alpha)t^2)$ with $m(\cdot,\cdot)$ a function of only $\theta_i$ and $2(1-\alpha)$. Therefore,
\begin{small}
	\begin{align}
		\label{DCT}
		D^i\mathcal{E}_\alpha(b ^k_i)&=\lim_{t\to 0}
		\Big[\int_{\boldsymbol{\Theta}}\psi_f(x,\boldsymbol{\theta})^{2\alpha}\prod_{j\neq i}(\psi_{q_j}(\theta_j))^{2(1-\alpha)}
		\Big\{2(\alpha-1)\psi^{1-2\alpha}_{q_i}(\theta_i)b^k_i(\theta_i)+\frac{1}{t}R(\theta_i,\alpha,t)\Big\}d\boldsymbol{\theta}\Big] \nonumber\\
		&=2(\alpha-1)\int_{\boldsymbol{\Theta}}\psi_f(x,\boldsymbol{\theta})^{2\alpha}\prod_{j\neq i}(\psi_{q_j}(\theta_j))^{2(1-\alpha)}\psi^{1-2\alpha}_{q_i}(\theta_i)b^k_i(\theta_i)d\boldsymbol{\theta}\nonumber\\
		&\qquad+\lim_{t\to 0}\int_{\boldsymbol{\Theta}}\psi_f(x,\boldsymbol{\theta})^{2\alpha}\prod_{j\neq i}(\psi_{q_j}(\theta_j))^{2(1-\alpha)}
		\frac{1}{t}R(\theta_i,\alpha,t)\Big\}d\boldsymbol{\theta}.
	\end{align}
\end{small}
For fixed $(x,\alpha) \in \mathcal{X} \times (0,\infty)\backslash \{1\}$, the sequence of functions
$$\boldsymbol{\Theta}\ni \boldsymbol{\theta}\mapsto H_t(x,\boldsymbol{\theta},\alpha):=\psi_f(x,\boldsymbol{\theta})^{2\alpha}\prod_{j\neq i}(\psi_{q_j}(\theta_j))^{2(1-\alpha)}\frac{1}{t}R(\theta_i,\alpha,t)$$ converges to $H(x,\boldsymbol{\theta},\alpha):=\psi_f(x,\boldsymbol{\theta})^{2\alpha}\prod_{j\neq i}(\psi_{q_j}(\theta_j))^{2(1-\alpha)}$ as $t \to 0$ with $|H(x,\boldsymbol{\theta},\alpha)|< \infty$ since $q(\boldsymbol{\theta})$ is strictly positive based on our assumptions. The result follows by an application of the dominated convergence theorem to the second integral in Equation \ref{DCT}.\\

\noindent\textbf{Proof of Proposition 4.}

We prove the result for the case when $\alpha=1/2$ to avoid cumbersome notation; the case of a general $\alpha$ can be worked out along the lines of the proof of Proposition 3 using the binomial expansion with real-valued powers. We use the result provided in Proposition 3.2 of \cite{RW} to prove our claim. They state that a sufficient condition to ensure the existence such an $\epsilon$ is that the restriction of $\tilde{\mathcal{E}}^{i}_\alpha$ to the linear span of the set $\{v_i\}$ be continuously differentiable.

Suppose we choose basis functions $\mathcal{B}_i$ for $T_{\psi_{q_i}}(\Psi)$ that are bounded (in $\mathbb{L}^2$) and smooth (e.g., Fourier basis). Observe then that from the expression of the gradient $\nabla \mathcal{E}^i_\alpha$ in Proposition 3, differentiability of $\tilde{\mathcal{E}}^{i}_\alpha$ depends solely on the existence and continuity of the directional derivative at $v_i \in T_{\psi_{q_i}}(\Psi)$ along a direction $b_i$, defined as
$$D^i\mathcal{E}_\alpha\circ\exp(v_i)(b_i):=\lim_{t \to 0}\frac{1}{t}\Big[\tilde{\mathcal{E}}_{\alpha|\Psi}(\psi_{q_i}+tb_i)-\tilde{\mathcal{E}}_{\alpha|\Psi}(\psi_{q_i}) \Big].$$
For convenience, denoting a function $h(\cdot)$ as $h$, similar calculations as in the proof of Proposition 3 results in
\begin{small}
	\[
	D^i\mathcal{E}_\alpha\circ\exp(v_i)(b_i)=\int_{\boldsymbol{\Theta}}\frac{\psi_f}{2 \text{exp}_{\psi_{q_i}}(v_i)}\frac{A(v_i)}{\|v_i\|}\Bigg[\frac{v_i}{2\|v_i\|^2}-\frac{\psi_{q_i}\sin(\|v_i\|)}{2}+\frac{b_i\sin(\|v_i\|)}{A(v_i)}+\frac{v_i\cos(\|v_i\|)}{2\|v_i\|} \Bigg]d\boldsymbol{\theta} ,
	\]
\end{small}
where $A(v_i):=\|b_i\|^2+2\|v_i\|\|b_i\|$. It's easy to verify that $D^i\mathcal{E}_\alpha\circ\exp(v_i)(b_i)$ is finite for every choice of $b_i$ and $v_i$ as long as $v_i \neq 0$.

The directional derivative $D^i\mathcal{E}_\alpha\circ\exp(v_i)(b_i)$ is clearly a linear operator from $T_{\psi_{q_i}}(\Psi) \to \mathbb{R}$. The space $\Psi$, which is the positive orthant of the unit sphere $S^\infty$ in $\mathbb{L}^2(\Theta_i)$, is an open subset of $S^\infty$ (Theorem 3.2 of \cite{KLMP}) and hence a Hilbert submanifold of $S^\infty$. The domain of the exponential map
$\text{exp}_{\psi_{q_i}}$ is all of $T_{\psi_{q_i}}(\Psi) $ and hence maps every element of $T_{\psi_{q_i}}(\Psi) $ to $\Psi$. Therefore it is a diffeomorphism from $T_{\psi_{q_i}}(\Psi) $ to $\Psi$ and clearly continuous.

Consider a sequence $v_{i,n}$ such that $\|v_{i,n} -v_i\| \to 0$. Noting that the norm function $x \mapsto \|x\|$ is continuous in $\mathbb{L}^2$, it is easy to see that every term inside the integral is a continuous function of $v_i$. This implies that $D^i\mathcal{E}_\alpha\circ\exp(v_{i,n})(b_i) \to D^i\mathcal{E}_\alpha\circ\exp(v_i)(b_i)$ as $n \to \infty$. For a finite $N$ chosen in the algorithm that determines the number of iterations, the argument can be extended to any linear combination of possible directions $\sum_{j=1}^N\alpha_{i,j}v_{i,j} $ on each tangent space $T_{\psi_{q_i}}(\Psi)$, and thus to the span under consideration. This completes the proof.

\section{Bayesian Linear Regression}

All of the following run time experiments were performed in Matlab on an Intel Core i7 processor (3.40 GHz) with 8 GB of RAM.

\subsection{Run Time Comparison for Gibbs Sampler and Proposed Method}

\begin{table}[!h]
	\centering
	\resizebox{\columnwidth}{!}{
		\begin{tabular}{|c|c|cc|cc|}
			\hline
			$d$   & $n$   & \begin{tabular}[c]{@{}c@{}}MSE Gibbs (\textit{Time})\\ \textit{iter}: 5000, \textit{burn-in}: 1000\end{tabular} & MSE PM (\textit{Time})        & \begin{tabular}[c]{@{}c@{}}MSE Gibbs (\textit{Time})\\ \textit{iter}: 30000, \textit{burn-in}: 5000\end{tabular} & MSE PM (\textit{Time})        \\ \hline
			25  & 100 & 1.0034e-05 (\textit{6.9151}) & 9.5201e-06 (\textit{1.3692})  & 7.0139e-07 (\textit{40.8964}) & 5.2551e-07 (\textit{1.8097})  \\ \hline
			50  & 100 & 1.7488e-05 (\textit{16.2920}) & 1.7234e-05 (\textit{9.9007})  & 4.9125e-06 (\textit{96.8816}) & 4.7913e-06 (\textit{11.2436}) \\ \hline
			75  & 100 & 2.2089e-04 (\textit{27.9082}) & 2.1717e-04 (\textit{24.2695}) & 1.8478e-05 (\textit{165.5516}) & 1.8268e-05 (\textit{39.1568}) \\ \hline
			100 & 200 & 7.0274e-06 (\textit{60.0844}) & 6.8176e-06 (\textit{37.2117}) & 1.0505e-06 (\textit{381.3203}) & 9.9089e-07 (\textit{47.4571}) \\ \hline
			
		\end{tabular}
	}
	\caption{Comparison of running times (in seconds) between the Gibbs sampler and the proposed method (PM) with $\alpha = 0.5$.}
	\label{tab:timeCompare}
\end{table}

In Table \ref{tab:timeCompare}, we compare the running times of our method to the Gibbs sampling approach. For each choice of $d$, a dataset of size $n$ is generated. First, we apply the Gibbs sampler with two choices of the total number of iterations and burn-in. The MSE in each case is noted along with the respective running time. Next, we fix $\alpha = 0.5$ and apply our method to the same dataset. We continue to update the corresponding $q_i$s in each case and run the algorithm as long as the MSE for the proposed method is larger than the one obtained using the Gibbs sampling technique. We stop the algorithm as soon as the MSE for our method becomes smaller than the Gibbs MSE and note the running time. It can be easily seen that the proposed variational method attains the MSE obtained by Gibbs sampling much faster in each case.

\subsection{Effect of Basis Size on Run Time}

\begin{table}[!h]
	\centering
	\begin{tabular}{|c|c|c|c|c|c|}
		\hline
		\multirow{2}{*}{} & $d = 10$  & $d = 25$  & $d = 50$  & $d = 100$ & $d = 200$ \\ \cline{2-6}
		& $n = 100$ & $n = 100$ & $n = 100$ & $n = 500$ & $n = 500$ \\ \hline
		$N = 49$           & 0.1151  & 0.5717  & 4.2978  & 12.6686 & 76.2938 \\ \hline
		$N = 99$            & 0.1748  & 0.9445  & 6.3781  & 14.1433 & 106.5231 \\ \hline
		$N = 199$           & 0.2892  & 1.4250  & 8.8313  & 21.6562 & 138.5783 \\ \hline
		$N = 499$           & 1.5809  & 7.7542  & 39.7014 & 72.8446 & 377.8602  \\ \hline
	\end{tabular}
	\caption{Comparison of running times (in seconds) for varying number of basis elements ($N$), sample size ($n$) and dimensionality ($d$) with $\alpha = 0.5$.}
	\label{tab:basistime}
\end{table}

In our setup, the tangent space for each $q_i$ is spanned by a finite collection of pre-specified orthonormal basis functions. We indicated in the main paper that increasing the number of basis elements leads to better approximations of the posterior. Table \ref{tab:basistime} shows the running times for our method for different choices of dimensionality $d$ and sample size $n$ as we vary the number of basis functions $N$. In this simulation, we use $\alpha = 0.5$. The proposed algorithm is fast for a moderately sized dataset even with $N = 199$ basis elements, i.e., run time is under 10 seconds for a dataset of dimension $d = 50$, and sample size $n = 100$.

\section{Bayesian Density Estimation}

\subsection{Effect of Sample Size on Density Estimation}

\begin{figure}[!h]
	\centering
	\resizebox{\columnwidth}{!}{
		\begin{tabular}{|c|c|c|}
			\hline
			& & \\
			\LARGE{$n = 25$} & \LARGE{$n = 200$} & \LARGE{$n = 1000$} \\
			\includegraphics[width=\columnwidth]{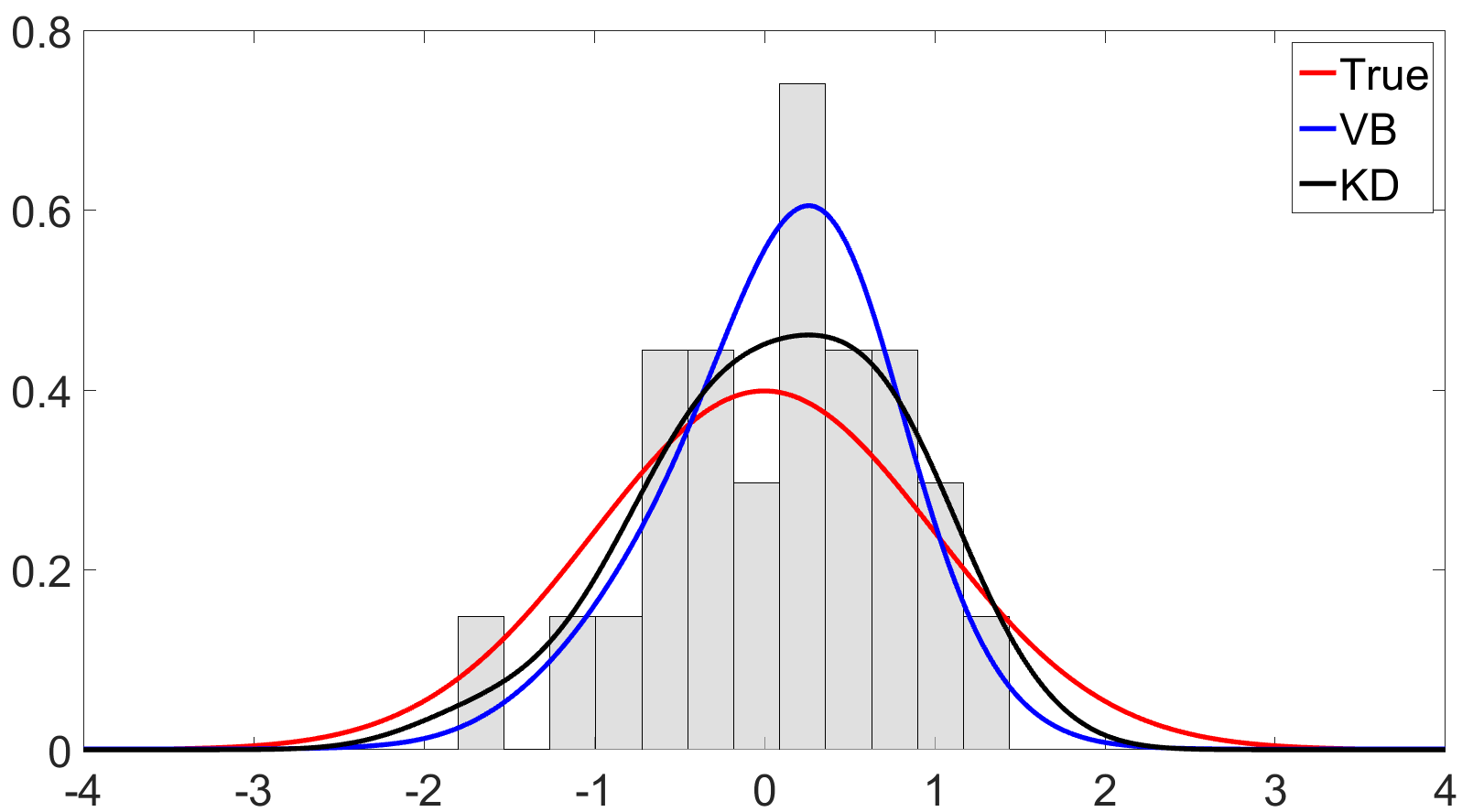} & \includegraphics[width=\columnwidth]{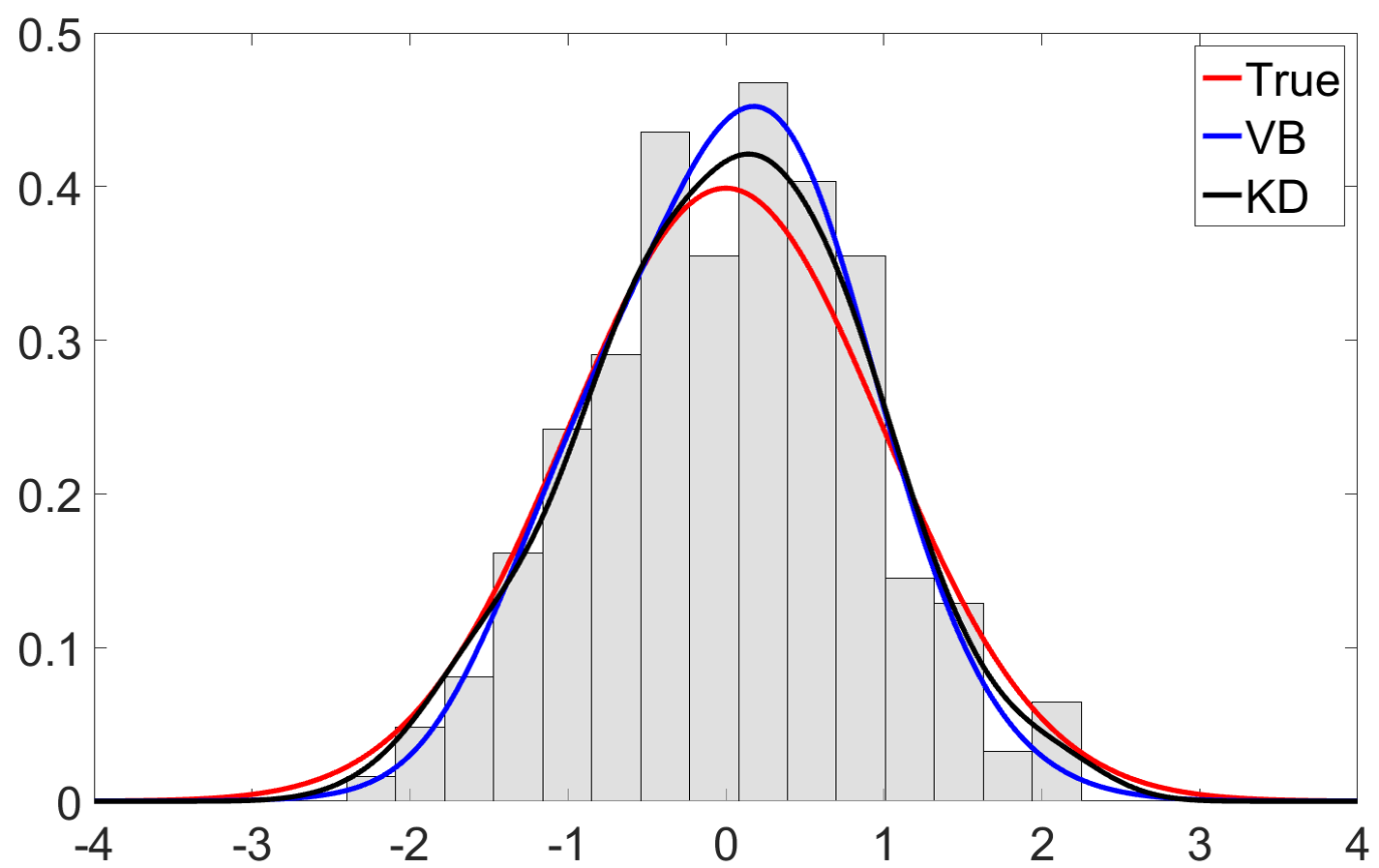} &
			\includegraphics[width=\columnwidth]{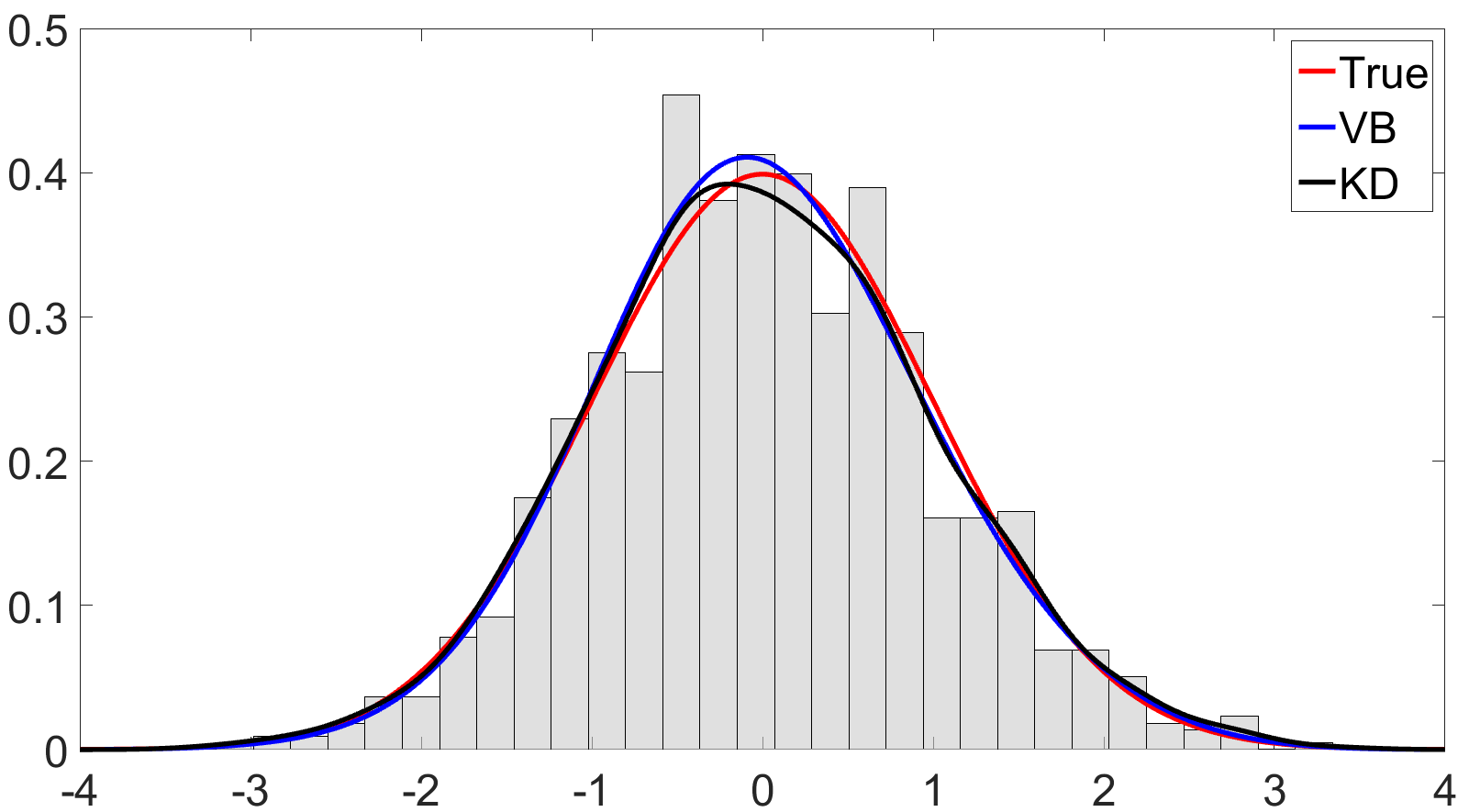} \\
			\hline
		\end{tabular}
	}
	\caption{Effect of increasing the sample size on density estimation. Data was generated from a N(0,1).}
	\label{fig:effect of sample size}
\end{figure}

Figure \ref{fig:effect of sample size} shows the effect of the sample size $n$ on density estimation under the proposed approach. As is common with other nonparametric density estimation techniques, we can see that with the increase in sample size, we obtain a better estimate of the true density function. Again, the proposed method with $\alpha=0.9$ performs favorably compared to standard kernel density estimation.

\subsection{Effect of $\alpha$ on Density Estimation}

\begin{figure}[!h]
	\centering
	\resizebox{0.95\columnwidth}{!}{
		\begin{tabular}{|c|c|}
			\hline
			\includegraphics[width=\columnwidth]{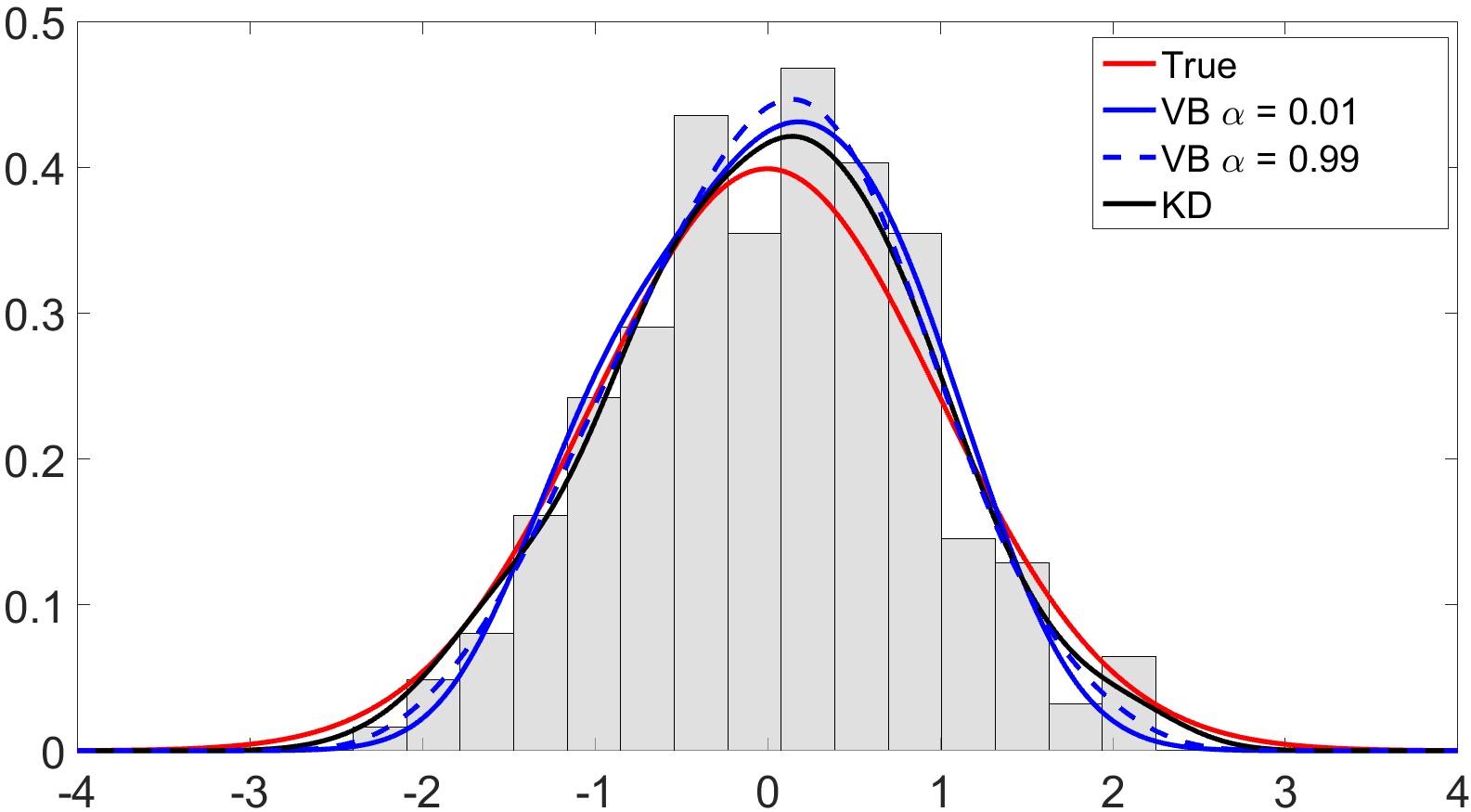} & \includegraphics[width=\columnwidth]{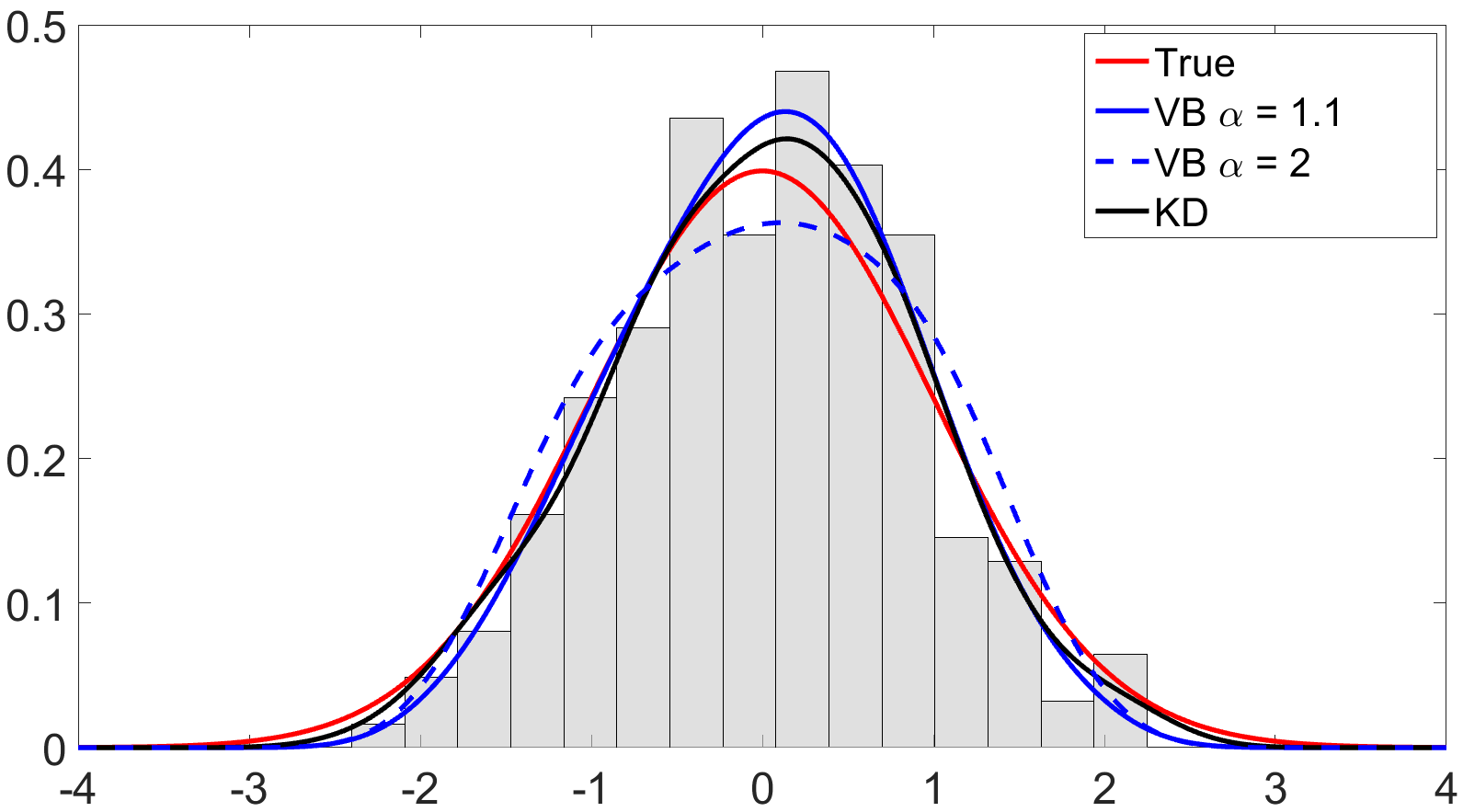} \\
			\hline
		\end{tabular}
	}
	\caption{Effect of different choices of $\alpha$ on density estimation. Data was generated from a N(0,1).}
	\label{fig:effect of alpha}
\end{figure}

Next, we assess the effect of $\alpha$ on density estimation in Figure \ref{fig:effect of alpha}. The plot in the left panel of Figure \ref{fig:effect of alpha} shows the results for standard  VB (i.e., $\alpha \rightarrow 0$) and EP (i.e., $\alpha \rightarrow 1$) in the limiting case, as noted in Section 3.1 in the main paper. We note that, using the proposed $D_\alpha$-based approach, we are able to explore a richer class of divergences with minor adjustments by simply changing the value of $\alpha$ accordingly. The right panel shows similar results for $\alpha=1.1$ and $\alpha=2$.

\section{Bayesian Logistic Regression for Real Data Applications}

We report the accuracy and average log predictive likelihood (ALPL) of the proposed method on two multi-label datasets \footnote{\url{http://mulan.sourceforge.net/datasets.html}} considered in \cite{wang2013}. As in the main paper MAP, PMEA, PMED and PPRED represent the maximum a posteriori, posterior mean, posterior median and posterior predictive, respectively, under the proposed approach. The results based on these summaries for KLD-based VB are also reported under KLMAP, KLMEA, KLMED and KLPRED.

\subsection{Yeast Data}
\label{subsub:yeast}

\begin{table}[!h]
	\centering
	\resizebox{\columnwidth}{!}{
		\begin{tabular}{c|cccc|cccc}
			\hlineB{3}
			& MAP  & PMEA & PMED & PPRED & KLMAP  & KLPMEA & KLPMED & KLPPRED  \\ \hlineB{1}
			Accuracy (in \%) & 79.5 & 79.5 & 79.4 & 79.6 & 79.3 & 79.2 & 79.3 & 79.2 \\
			ALPL & -0.6587 & -0.6579 & -0.6595 & -0.6576 & -0.6718 & -0.6731 & -0.6734 & -0.6723 \\
			\hlineB{3}
		\end{tabular}
	}
	\caption{Classification results for the the yeast dataset.}
	\label{tab:yeast}
\end{table}

The yeast dataset \citep{elisseeff2001} is composed of micro-array expression data and phylogenetic profiles with 1500 genes in the training set and 917 in the test set. For each gene, we have 103 covariates and up to 14 different gene functional classes, making it a multi-label problem. This can be reformulated into 14 separate binary classification tasks, similar to the problem studied in \cite{wang2013}.

We use 299 basis elements for this dataset, and choose $\alpha = 0.9$. Table \ref{tab:yeast} presents the classification results, averaged over the 14 binary problems, obtained under this setup. The results for all of the methods are very similar in this case. The ALPL values based on all of the $D_\alpha$ summaries in Table \ref{tab:yeast} are greater than the KLD-based summaries, indicating better fit. \cite{wang2013} report accuracy rates of 80.1\% and 80.2\%, and ALPL values of -0.449 and -0.450 using Laplace inference and delta method inference, respectively. The method of \cite{jaakkola1997} gives accuracy of $79.7\%$ and an ALPL value of $-0.678$. However, note that both of those papers used cross-validation to evaluate the performance of their algorithm, which is different from the training-testing split considered here. Nonetheless, the proposed method produces results that are comparable to the previously reported classification results.

\subsection{Scene Data}
\label{subsub:scene}

The scene dataset was used for the problem of semantic scene classification in \cite{boutell2004}, where a scene might contain different objects such that it can be described by multiple class labels. The dataset contains 1,211 images in the training set and 1,196 images in the test set. The images consist of 294 image features that can be used to predict scene labels. There are up to six scene labels per image. Analogous to the previous example, this corresponds to six separate binary classification problems.

We use 299 basis elements to estimate the energy gradient in this classification problem, and choose $\alpha = 1.1$. The classification results, averaged over the six binary problems, obtained using the summary measures based on the proposed $D_\alpha$-based VB and KLD-based VB are presented in Table \ref{tab:scene}. With the exception of the posterior median, summaries based on $D_\alpha$ perform better than those of \cite{jaakkola1997} (they achieve $87.4\%$ accuracy). Moreover, the ALPL values based on $D_\alpha$ summaries are significantly higher as well (they obtain a an ALPL value of $-0.670$). Our classification accuracy results are comparable to the results obtained using Laplace inference and delta method inference introduced in \cite{wang2013}. Similar to the yeast example considered in Section \ref{subsub:yeast}, the results in those papers were generated via cross-validation, which is different from our training-testing split setup.

\begin{table}[!t]
	\centering
	\resizebox{\columnwidth}{!}{
		\begin{tabular}{c|cccc|cccc}
			\hlineB{3}
			& MAP  & PMEA & PMED & PPRED & KLMAP  & KLPMEA & KLPMED & KLPPRED  \\ \hlineB{1}
			Accuracy (in \%) & 88.3 & 88.6 & 84.1 & 88.6 & 82.7 & 85.7 & 83.1 & 85.9 \\
			ALPL & -0.5649 & -0.5596 & -0.6375 & -0.5584 & -0.8375 & -0.7653 & -0.8425 & -0.7581 \\
			\hlineB{3}
		\end{tabular}
	}
	\caption{Classification results for the scene dataset.}
	\label{tab:scene}
\end{table}

\newpage
\bibliographystyle{Chicago}
\bibliography{References}

\end{document}